\documentclass[twocolumn,english,superscriptaddress,floatfix]{revtex4-2}
\usepackage[T1]{fontenc}
\usepackage[latin9]{inputenc}
\usepackage{array}
\usepackage{textcomp}
\usepackage{graphicx}
\usepackage{multirow}

\makeatletter 

\providecommand{\tabularnewline}{\\}

\makeatother

\usepackage{babel}
\begin{document}
\title{Physicality, Modeling and Agency in a Computational Physics Class}
\author{A. M. Phillips}
\affiliation{Tufts University Department of Physics \& Astronomy, 574 Boston Ave,
Medford, MA 02155}
\author{E. J. Gouvea}
\affiliation{Tufts University Department of Education, Paige Hall, 12 Upper Campus
Road, Medford, MA 02155}
\author{B. E. Gravel}
\affiliation{Tufts University Department of Education, Paige Hall, 12 Upper Campus
Road, Medford, MA 02155}
\author{P.-H. Beauchemin}
\affiliation{Tufts University Department of Physics \& Astronomy, 574 Boston Ave,
Medford, MA 02155}
\author{T. J. Atherton}
\affiliation{Tufts University Department of Physics \& Astronomy, 574 Boston Ave,
Medford, MA 02155}

\begin{abstract}
Computation is intertwined with essentially all aspects of physics research and is invaluable for physicists' careers. Despite its disciplinary importance, integration of computation into physics education remains a challenge and, moreover, has tended to be constructed narrowly as a route to solving physics problems. Here, we broaden Physics Education Research's conception of computation by constructing an epistemic \emph{metamodel}---a model of modeling---incorporating insights on computational modeling from the philosophy of science and prior work. The metamodel is formulated in terms of practices, things physicists do, and how these inform one another. We operationalize this metamodel in an educational environment that incorporates making, the creation of shared physical and digital artifacts, intended to promote students' agency, creativity and self-expression alongside doing physics. We present a content analysis 
of student work from initial implementations of this approach to illustrate the very complex epistemic maneuvers students make as they engaged in computational modeling. We demonstrate how our metamodel can be used to understand student practices, and conclude with implications of the metamodel for instruction and future research.
\end{abstract}

\maketitle

\section{Introduction}

Despite its fundamental importance to physics practice, integration of computation into physics pedagogy has only begun to take place surprisingly recently. Responding to the difficulty of sustaining educational change in higher education\citep{Dancy.2010,Dancy.2016}, initiatives such as the Partnership for Integration of Computation into Undergraduate Physics (PICUP) have successfully built supportive communities for faculty to develop and adopt computational projects and assignments\citep{Caballero.2019,Behringer.2017} and the American Association of Physics Teachers (AAPT) has produced a valuable report to facilitate institutional changes\citep{behringer2017aapt,Behringer.2017}. While these efforts have met with success, large-scale adoption of modern computation in physics education and integration with interactive engagement methods remains challenging\citep{Martin.2016,Caballero.2018}. Where opportunities for students to engage in computation remain limited or unavailable, students are unable to learn these ever more important skills. 

One of the main objectives of efforts to promote computation is to better prepare students for active contributions to physics 
research\citep{Graves.2019} or for contemporary careers in STEM\citep{McNeil.2017}. The key to such preparation is to 
offer students the opportunity to engage a broad range of professional scientific practices in the classroom\citep{Chabay.2008,Weber.20200ku}. 
However, computational projects and assignments in Physics Education are often framed as problems on particular domains of physics involving systems specified by the instructor\citep{lane2021analysis}. A trained scientist, however, is not someone who has simply studied a large set of such exemplars: they have also developed a critical evaluation of the problems to be able to use them as resources,  with appropriate justification, for solving novel problems or conducting new scientific inquiries. Hence, critical evaluation must be central to scientific training.

An alternative framing that better reflects how physicists use computation would instead focus on computational \emph{practices}, such as numerical analysis, coding, testing, and visualization\citep{Caballero.2015,Burke.2017,behringer2017aapt}. Such a framing lends itself naturally to a project-based approach as has been developed by several authors\citep{Rebbi.2008,Burke.2017,Caballero.2018,Langbeheim.2020sg}, and provides us with an opportunity to utilize work in K-12 science education scientific practice \cite{ford2008grasp, berland2016epistemologies, aleixandre2017epistemic}. As with the National Research Council's Framework on K-12 science education, we ``use the term `practices' instead of a term such as `skills' to emphasize that engaging in scientific investigation requires not only skill but also knowledge that is specific to each practice'' \citep[p. 30]{national2012framework}. Understanding computational physics practices in classrooms and professional physics is valuable to physics education broadly since computation is inextricably interwoven into all aspects of physics: Computers are used to design experiments, collect, analyze and visualize data, perform simulations and construct and test mathematical theories. 

Motivated by this enriched perspective on computation,
it is natural to ask: \emph{How do we design educational environments
where students can engage in computational practices and other physics practices
and how can we theoretically ground such design
?} As an answer this question, in this paper we will propose a \emph{metamodel}---a model of modeling processes---of knowledge production in computational physics that will then be leveraged to design and analyze projects intended to empower students to cultivate these practices. Our work here is inspired by our reading of relevant literature from philosophers of science: Kent Staley provides an epistemological perspective on experimentation grounded in scientific practice\cite{staley2004robust,staley2012strategies,staley2020securing} while Paul Humphreys offers a nuanced perspective on simulation and more generally on computational modeling\citep{humphreys2004extending}. We therefore define computation as much more than just a set of methods used to numerically solve problems that cannot be solved analytically but as a dynamic set of \emph{practices} that are central and necessary to the production of scientific knowledge, and that are explicitly or implicitly performed by scientists when they successfully create physics knowledge. 

Note that our use of the term ``knowledge'' deliberately does not commit to a particular conceptual analysis of that idea. Following Dewey\citep{deweylogic} and Pierce\citep{peirce1992essential}, we instead draw from the pragmatic tradition in philosophy of science to think of knowledge as the product of a successfully executed act of inquiry, where its success is evaluated by the acceptability of the outcome to a community of investigators. A pragmatic account of computation therefore seeks to describe how computational activities, suitably conducted, generate outcomes that can serve as the premise of future inquiries, including experimental ones, but could also be further refined. 

By framing our approach to computation in pragmatic philosophy, we can adopt a working conception of model, theory, and knowledge that evades contested epistemological analyses. Rather, we can focus on the development of teaching strategies that rely on the expertise of practicing scientists who are not trained in epistemology. The pragmatic approach aligns well with PER efforts to re-position the classroom as a place for students to \emph{do physics} rather than imbibe knowledge\citep{hammer1996more}. It also deters us from evaluating student work as correct or incorrect, in line with PER recommendations\citep{hammer1996more, Hammer.2000}. The emphasis of our study will therefore be on what is done by students, the things they produce, the resources they use and the relationships between practices, resources and products, rather than on the syntactic or semantic structure of computation. As we will discuss in Sec. \ref{sec:modeling}, this is a very different perspective on computational modeling than has been previously adopted in PER.



As will be discussed in subsequent sections of this paper, our new metamodel provides a theoretical foundation for education design and research in computational physics. It is used to ground our design research\citep{brown1992design} approach to explore how students engage in, and learn with, computational practices.  Design research allows us to build new educational environments while also developing new theoretical positions on how learning takes form in computational activity. Motivated by our metamodel, we adopt \emph{making}, an open-ended but intentional learning activity that centralizes self-expression, materiality, design, and iteration\citep{halverson2014maker,tucker2019stem}. As we shall show, making enables students to produce and computationally model phenomena in a way that reflects physics disciplinary practices that are essentially absent in typical classrooms. 

In Section \ref{sec:Background} of this paper, we examine prior research on modeling and metamodeling in science education research, and we provide an expanded view of computational modeling from philosophy of science. We then develop, in Section \ref{sec:EpistemicModel}, our new metamodel for computational physics taking into account these considerations. Note that our metamodel is not limited to computational physics contexts but can be used in a broader range of learning environments. In Section \ref{sec:Design}, we describe the design of the course projects which operationalized our metamodel, while the details of their implementation are presented in Section \ref{sec:Implementation}. Using data collected in Spring 2019, we examine student work in Section \ref{sec:Student-Work} using our metamodel to identify how students engaged in computational modeling within the designed context. We discuss these initial experiences in Section \ref{sec:Discussion} and compare our approach with other theoretical perspectives on computation in physics. We also propose possible research directions to understand how to further optimize these practices, and conclude with the implications of our metamodel in Section \ref{sec:Conclusion}. 

\section{Background}\label{sec:Background}

\subsection{Modeling and Metamodeling In Physics Education}
\label{sec:modeling}

Metamodels have proven a productive resource for Physics Education Researchers to design and assess interventions, from curricula to individual assignments, from the foundation of the field. In pioneering work, Hestenes proposed a model-centered instructional strategy that focused on building students' conceptual understanding of physical reality\citep{Hestenes.1987}. Nonetheless this early work did not consider computational models specifically, and philosophical understanding of modeling has evolved significantly since then (see for example \citep{humphreys2004extending, morrison2015reconstructing, gelfert2016science}, to name a few). In addition, Hestenes focused on the structure of scientific knowledge he saw in professional scientific practices, rather than on the knowledge creation and revision processes at play in these practices, limiting the practical use of such metamodeling in physics education. These limitations motivate the present work. 

Nevertheless, the resulting Modeling Instruction Curriculum met with considerable success and has inspired a number of related efforts. Of particular pertinence to the present work is the Matter \& Interactions curriculum that includes a significant computational component\citep{Chabay.2008} and popularized the VPython environment within the physics community. Chabay \emph{et al.} \citep{Chabay.2008} describe the benefits of computation in this context as providing new representations through data structures and visualization as well as promoting open-ended exploration and stimulating creativity. 

Brewe's careful re-exposition of Modeling Instruction applied to introductory physics \citep{Brewe.2008} presents a cyclic metamodel that includes as steps identification of phenomena and appropriate representations, coordination of representations, application of knowledge and tools, abstraction and generalization and continued incremental development. Like Hestenes, Brewe emphasizes the reflection of scientific practice in pedagogy and notes the rich problem-solving experiences that this makes possible.

A number of authors have since developed more sophisticated metamodels. Zwickl \emph{et al.} \citep{Zwickl.2014lgj} constructed a metamodel for modeling in physics laboratories that incorporates both modeling of the physical system of interest as well as the measurement process, comparison of the results with predictions and iteration. In a later paper\citep{Zwickl.2015}, they clarified their metamodel slightly to emphasize that iteration could include revision of the physical or measurement model or the experiment. Their metamodel provided a rich tool to analyze student thinking in think-aloud interviews based on their lab work\citep{Zwickl.2015}. 

As a vivid illustration of the utility of metamodels as theoretical tools for design and analysis in PER, a number of other authors have used, and even further refined, Zwickl {\it et al.}'s metamodel, later referred to as the \emph{Modeling Framework for Experimental Physics}\citep{dounasfrazer2018}, to examine the intersection of modeling and troubleshooting \citep{Dounas-Frazer.2016} or to develop assessment instruments for electronics laboratories\citep{Rios.2019}, to cite a few examples. By carefully analyzing video data of student work, references \citep{Dounas-Frazer.2016,Rios.2019} reveal that students engaged in different modeling subtasks almost simultaneously, which parallels some of our earlier observations about the deep interconnection between various modeling components as we reported in \citep{Burke.2017}.

A serious limitation of the above metamodels developed for Physics Education is that they conceive modeling solely in its representational relationship to phenomena. However, models could also be used, for example, to predict, design protocols, enhance performance, develop critical evaluation of other models or of themselves, build something, or even price a product. In a valuable recent paper, Russ and Odden draw upon research from psychology and philosophy of science to argue that modeling should indeed not be separated from evidence based reasoning, because both these activities are located in a broader epistemological context\citep{Russ.2017}: modeling guides which evidence to collect, to create mechanistic explanations, and to make claims about causality. 

\begin{figure}
\includegraphics{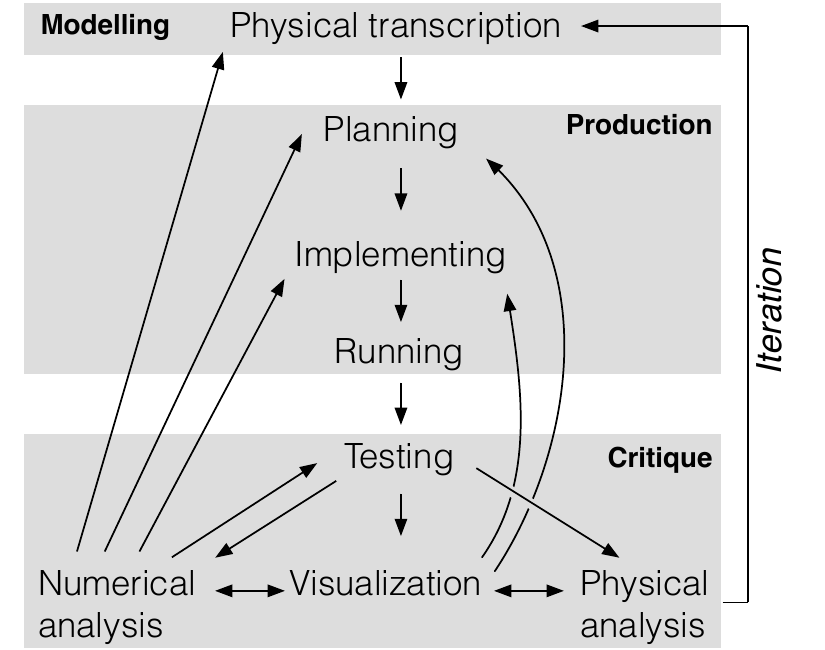}

\caption{\textbf{\label{fig:Prior-epistemic-model}Prior metamodel} of knowledge production in computational physics\citep{Burke.2017}. Arrows denote how the results of certain practices are used by a scientist to facilitate others. 
}
\end{figure}

In a prior paper\citep{Burke.2017}, we presented an earlier iteration of our computational physics course designed to encourage students to engage in scientific practices. Central to the design was a tentative metamodel, depicted in Fig. \ref{fig:Prior-epistemic-model}, of computational physics that was constructed from literature and interviews with domain practitioners. This model proved useful for connecting the practices of professional physicists to those of our students as well as designing rubrics: \citep{Burke.2017} contains detailed descriptions of what these practices might look like in student work. However, as we continued to develop the computational physics course, we found this metamodel did not capture the non-linear and complex practices we saw in our students and our own work as physicists who use computation in research. These observations motivated us to develop the revised metamodel presented in Section \ref{sec:EpistemicModel}.


Some practices in our prior metamodel were also identified as important in other physics education work\citep{Caballero.2015,behringer2017aapt}, and its iterative nature parallels other models that have been used to analyze computation in science and engineering\citep{litzinger2006cognitive, litzinger2010cognitive, vieira2015exploring}. Nonetheless, the connectivity that we proposed departs significantly from prior metamodels explicitly or implicitly proposed in PER to study computation: we found that scientists tended to hop between practices in a very complex manner rather than in a strict order. This parallels observations in other disciplines\citep{schoenfeld1987s, shoenfeld2013mathematics, dym1994engineering}. Further, by focusing on the \emph{practices} involved in an inquiry, rather than the content of a model, our prior metamodel is not limited only to representational modeling and is already implicitly grounded in pragmatic philosophy. 

To redesign our metamodel, we need a better understanding of what modeling phenomena entails. In the next section draw upon impressive philosophy of science work\citep{humphreys2004extending,staley2004robust,staley2012strategies,staley2020securing} that provides a sophisticated view of computational models. Note that while the question of what constitutes a model is certainly very complex (see for example \citep{morrison2015reconstructing}), adopting a pragmatic approach allows us to focus instead on how physicists engage in modeling practices in the production of computational physics knowledge. 

\subsection{Computational Models}\label{sub:compmodels}

Perhaps surprisingly, given the connections between computation and modeling, the question of how \emph{computational models} might differ from other kinds of model---and the consequences for instruction---has received little scrutiny within PER. Fortunately, the work of the philosopher Paul Humphreys\citep{humphreys2004extending} offers sharp insight into this question. Humphreys argues that computational models have six components: 

\begin{enumerate}
\item A \emph{computational} \emph{template}, which incorporates both the equations to be solved and additional necessary information specified. For example, the Schr\"odinger equation or Newton's second law are schemas or \emph{theoretical templates} that are computationally intractable until a potential or force function is specified. We note that the choice of these additional components themselves can be models, and may or may not be computational in nature. Moreover, computational templates need not be built on laws or theories (as in agent-based modeling and some types of Machine Learning). 

\item \emph{Construction assumptions}. These can include both the assumptions of the theoretical templates used and those adopted specifically for the model and not drawn from a general theory. Assumptions might include the type of model required, and what abstractions, constraints and approximations are appropriate. Some of these assumptions might be for physical reasons, but others could be for pragmatic reasons, i.e. to facilitate calculations. 

\item The \emph{correction set} which specifies in advance the ways in which the computational template must be adjusted if it fails to match empirical data. This necessarily involves changes to the construction assumptions, such as refining approximations or relaxing constraints. 

\item An \emph{interpretation} of how the model is supposed to represent the physical system, incorporating both ontology (a map from components of the model to components of the physical system) and causal structure. 

\item The \emph{initial justification} of the template which may be revised for empirical or pragmatic reasons. 

\item An \emph{output representation}, including a data file or visualization. 
\end{enumerate}

As a very brief example, a scientist might construct a computational model of a pendulum using Lagrangian mechanics together with a gravitational potential and an Euler-Cromer integrator as the \emph{computational} template. The scientist might initially model the real pendulum as a point mass connected to a massless inextensible string \emph{in vacuo}---the \emph{construction assumptions}. The model's \emph{correction set}, ways that the model could be revised, might include incorporating the mass of the string, drag, etc. or revising the choice of integrator. The model's \emph{interpretation} is that the trajectory of the mass is explicitly represented by a (time dependent) position vector, while the string is represented by a constraint. \emph{Justification} of the template used might involve prior knowledge of the physical string's mass relative to the bob. The completed program might produce a number of \emph{output representations}, including the raw trajectory data from the program's output, graphs of the trajectory or even an animation. 

Even for such a simple physical system, Humphreys' picture enables us to see the many elements of a computational model that are not simply representational. It also describes how a model may be revised: Once the initial template is chosen, for example, it can be refined by assessing the assumptions, comparing with data and adjusting the template according to the correction set. The initially predicted trajectory of the pendulum may not match the experimental data, for example, and show a diminishing amplitude; the scientist might therefore suspect the need to incorporate damping into the model. Hence, implicit in this account of what a computational model \emph{is} is a metamodel of how such models are to be constructed and revised that is described in detail in \citep{humphreys2004extending}. 

Humphreys' conception of computational modeling focuses our attention on the epistemic import of models, helps us locate them in an epistemological framework and accounts for nuances in how they are used in science. It invites us to look beyond the representational aspects of a computational model. The richness of the above construction provides a detailed lens to compare models: The same computational template could be used to model different systems, even in different disciplines, and the differences between such models would lie in the correction set, justification or interpretation. Similarly, two models could have strongly overlapping assumptions and a common or nearly common correction set, but use different computational templates. Our example model of a pendulum might equally well have been constructed using Newton's laws rather than the Lagrangian approach, without one outcome being more `true' than the other.

Here we will adopt a complementary but distinct approach from Humphreys. He, as we have described, tried to identify the components of a computational model and showed that they are more than a theory or a representation and instead comprise a network of elements that enable a `core' theory to reach calculated predictions. Practitioners of science, however, need not have a clear understanding of what a model is, but are still able to make one. By focusing on \emph{practices}, what is done in computational modeling, rather than on the structure of a model, and treating models themselves as resources to be used to perform these activities, we recover many of the same components that Humphreys identifies, while being able to operationalize these components, i.e. to map them to what scientists are actually doing when they are modeling.

For the present work, Humphreys' work motivates us to think more expansively about what \emph{could} happen in a physics classroom and how to make students think more explicitly about their modeling work. Alongside creating programs, exploring algorithms and visualizing data, etc., students could also, to name just a few possibilities, justify their models, critically examine different assumptions, templates and even entirely different models, and embark on creating and revising models in response to data. In order to facilitate this, we will develop in the next Section a practices-based view of computation, in effect a revised version of Fig. \ref{fig:Prior-epistemic-model}, that incorporates some of Humphreys' key insights, but adapted to the pragmatic epistemological framework adopted here that allows us to more directly connect to the work and practice of students and physicists.

\section{Metamodel\label{sec:EpistemicModel}}

\begin{figure*}
\includegraphics[width=1\textwidth]{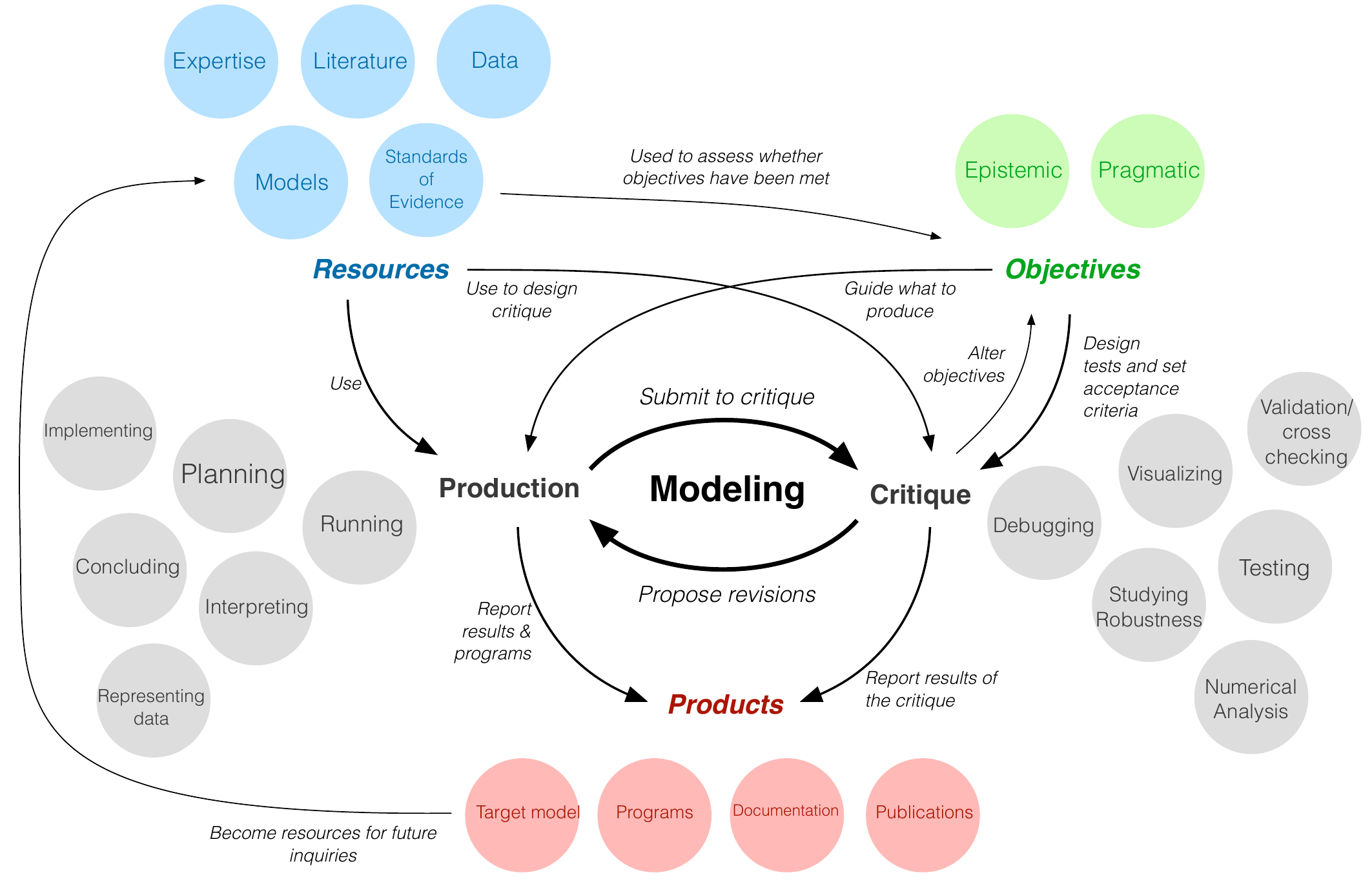}

\caption{\textbf{\label{fig:Revised-epistemic-model}Revised  metamodel} of knowledge production in computational physics including the following five components: In a scientific inquiry, a scientist undertakes a cycle of \emph{production} and \emph{critique} practices to meet their \emph{objectives}. In doing so, they use \emph{resources}. The outcome of a successful inquiry are scientific \emph{products} that become resources for future inquiries. Illustrative examples of each component are shown as disks. Arrows depict how elements of the metamodel inform or are used by one another in an inquiry.}
\end{figure*}

To construct our revised metamodel of computational physics, we begin by re-examining our prior metamodel: We observe that the web of practices in Fig.~\ref{fig:Prior-epistemic-model} may be considerably disentangled by dividing them into two categories, which we shall refer to as \textbf{production} and \textbf{critique} respectively. These practices inform, utilize, and create \textbf{resources}, \textbf{products}, and \textbf{objectives}. These resources, products and objectives consist of tangible artifacts such as programs, references, and reports as well as abstractions such as computational and conceptual models and knowledge. 

The new metamodel is displayed schematically in Fig. \ref{fig:Revised-epistemic-model} which depicts how the various components (boldface text) inform one another (arrows) together with examples of each component (disks). The examples enumerated in our revised metamodel are not meant to be comprehensive, but rather what we have identified through literature, as discussed above, as well as student and professional practice, as discussed in Ref. \cite{Burke.2017} and later in this paper. To assist the reader in interpreting our description of the metamodel, examples of the enumerated items from students engagement in computational physics are shown in tables \ref{tab:knowledge} and \ref{tab:activities}. We provide these here to assist in defining the various terms used. The context for these examples is provided in Sections \ref{sec:Design} and \ref{sec:Student-Work}.

Production practices relate to the creation of scientific artifacts and ultimately products; critique practices relate to the evaluation of the outcomes of the production tasks and the resulting computational model with respect to the objectives of the inquiry and the standards that are targeted. These include standards of evidence (e.g. estimates of uncertainties on quantitative results) which are used as resources to assess if the objectives of a scientific inquiry have been met, as discussed by Beauchemin in \citep{beauchemin2017autopsy}, and by Ritson and Staley in \citep{ritson2021uncertainty}. 

As illustrated in Fig. \ref{fig:Revised-epistemic-model}, practitioners draw upon a variety of resources to perform the scientific practices that enable them to meet the objectives of their inquiry. Resources include many kinds of existing model, theories and templates, data, experimental results, their own expertise, aforementioned standards of evidence etc. Some of these, such as literature, are artifacts that exist in tangible form. Others, such as expertise and models, may primarily exist as knowledge or other types of abstractions.  The same resources may be used for many different scientific inquiries, but some practices might also require very specific resources. Production and critique are therefore seen as two different ways of using resources to meet the epistemic objective(s) of the inquiry, but also at validating the computational model resulting from the entire process. We do not consider the resources, the products, or the practices that we listed under production and critique to be exhaustive. Professional physicists as well as students may engage in a broader range of practices that would fall under these umbrellas. Similarly, ``resources'' should be understood broadly. For example, we consider expertise to not only consist of the skills developed over years of practicing physics, but also scientists' expertise developed in their daily lives including ways of constructing, observing, and knowing. 

In this view, progress is made by the scientist by producing a computational artifact and data, while performing critique. The scientist will test whether the model yields internally consistent results, for example if the approximations made are \emph{a posteriori} found to be satisfied, and conserved quantities are indeed conserved. They will also determine whether the results satisfy their objectives. If not, they may revise components of the model or resort to further production. New resources might also be brought to bear. They may even revise the objectives themselves or modify the standards of evidence desired, e.g. by increasing or lowering the targeted precision of a calculation. New tests might then be proposed, the computational model revised or further critique performed and the process continues. Hence, ``iteration" in our new metamodel ultimately corresponds to revising the resources and/or objectives based on data. The process continues until the scientist meets the objectives of the inquiry with reference to standards of evidence---precision, statistical significance, size of dataset---that are acceptable to them and arising from their disciplinary context. The outcomes of the inquiry, which could include programs, manuscripts, data, presentations and new knowledge then become resources for future inquiries.

The above epistemological account is inspired by the epistemological framework proposed by Kent Staley for securing evidence produced in an experimental inquiry\citep{staley2004robust,staley2012strategies,staley2020securing} and is fully rooted in a pragmatist perspective. Staley and one of the authors of this paper (PHB) are in process of developing this framework into a broader epistemological model of experimentation in the physical sciences. Our metamodel proposed here represents an application of this work to computational problems. We therefore anticipate that our new metamodel, presented in Fig. \ref{fig:Revised-epistemic-model}, might be further revised in the future as the broader framework emerges. Our new metamodel nevertheless certainly already constitutes a very powerful tool to be used for educational purposes.

Our revised metamodel provides a design tool that enables us to identify instructional activity within this framework as well as a diagnostic tool to understand how students engaged in the various scientific practices in computational physics. It allows us to create educational materials that focus on some or all of the computational practices elucidated by our metamodel. For example it is possible to envisage a project primarily focused on the production mode, which would involve a concrete, well-posed problem and the students writing and running the code. Similarly, a project focused on the critique mode might involve running a pre-existing code or using an existing data set and having the students interpret the output. 


One of the key elements of our metamodel, which was also an important feature of our earlier work, is that scientists (including students as shown in ref. \citep{Dounas-Frazer.2016,Rios.2019,Russ.2017}) maneuver within these categories frequently and arbitrarily, and also traverse between them: for example, having produced data the scientist offers it for critique, which may then involve revisions to the products or the resource models. The iterative nature of the metamodel therefore allows for the possibility that the inquiry process can itself impact the resources that are used in the process, and even lead to the creation of new resources to advance the inquiry. This mirrors work in engineering education on the ``co-evolution'' of problems and solutions \cite{maher1996modeling, dorst2001creativity, watkins2014examining}, and work in philosophy of science describing how the natural world ``pushes back'' on our understandings to drive scientific advancement \cite{pickering2010mangle}.
The assumptions made, the level of sophistication of the models used to solve a particular computational problem, and the techniques employed in the production mode can all be modified as a result of critique (as Humphreys envisages). With reference to our example in Section \ref{sub:compmodels}, the scientist's desire to improve on the precision of the pendulum computational model might lead to using more sophisticated models of friction and mass distribution as input to the calculation being performed. 

The space of possibilities offered by all the resources that could be brought to bear is very large. The process by which the inquiry objectives are met and the products that result from the inquiry are not unique. As a consequence, different choice of resources can be used to reach the same ends, and there are multiple ways by which these ends could be reached.

Within the scope of an inquiry, some models may be  \emph{protected}, i.e. ones that the scientist is unwilling to revise, and other may be \emph{revisionable}, i.e. those that the scientist is prepared to modify as a consequence of critique practices. We stress that ``protected'' and ``revisionable'' models are not used here in the Lakatosian sense\citep{lakatos1976falsification}; what we consider as protected or not is specific to a \emph{particular computational problem}. The category of revisionable models approximately corresponds to Humphreys' notion of the correction set associated with a computational model, but is framed here from the viewpoint of practices consistent with the pragmatic approach adopted in this work. 

Protected resources typically consist of background knowledge about both target and ancillary phenomena, as well as knowledge of computational models that have been successful in the past for the relevant domain of physics. They are used to inform the production activity as well as constrain the design and implementation. Revisionable models will often include hypotheses about the target and ancillary phenomena, extension of background knowledge about both types, and new features of the computational model. Critique activities may lead the scientist to propose revisions of these components of the model. Both protected and revisionable modeling are used to design tests that facilitate critique: For example they are often used in estimates of systematic uncertainties. 

Our distinction between revisionable and protected 
accounts for the observation that there exists a multitude of pathways to solve a problem. Hence, a scientist will typically adopt a pragmatic attitude of limiting the space of resources that will be put into scrutiny at one time. The very same computational problem tackled by a different team of scientists can, and commonly does, lead to a different set of protected and revisionable models. Ultimately, no model is always protected over the space of all scientific activity, but in an individual investigation a particular model may be. Since the investigator has the choice of which resources will be used in the inquiry, as well as what will be protected and what will not, the same should apply to the classroom: students should be free of the resources they use to solve a given problem, and of deciding which ones can be revised and which ones will not. 

\begin{table*}
\begin{tabular}{|>{\centering}p{0.75in}|>{\raggedright}p{1in}|>{\raggedright}p{1.6in}|>{\raggedright}p{3in}|}
\hline 
 & \textbf{Construct} & \textbf{Definition} & \textbf{Example}\tabularnewline
\hline 
\multirow{2}{0.75in}{\textbf{Objectives}} & Epistemic Objectives & {\small{}The target knowledge} & {\small{}Half-pipe group initially seeks to model the motion of their
oscillator}\tabularnewline
\cline{2-4} \cline{3-4} \cline{4-4} 
 & Pragmatic Objectives & {\small{}Practical targets} & {\small{}Students aim to create code that runs efficiently}\tabularnewline
\hline 
\multirow{5}{0.75in}{\textbf{Resources}} & Expertise & {\small{}Experience and background knowledge} & {\small{}All students had at least 2 years of undergraduate physics
training.}\tabularnewline
\cline{2-4} \cline{3-4} \cline{4-4} 
 & Models & {\small{}An existing model} & {\small{}Multiple groups used the Lagrangian approach.}\tabularnewline
\cline{2-4} \cline{3-4} \cline{4-4} 
 & Data & {\small{}(Numerical) information gathered for the purpose of analysis} & {\small{}All groups collected video data.}\tabularnewline
\cline{2-4} \cline{3-4} \cline{4-4} 
 & Standards of Evidence & {\small{}Reference criteria to assess quality of evidence available} & {\small{}All groups used visual superposition of experimental and
simulated trajectory as a criterion for fit.}\tabularnewline
\cline{2-4} \cline{3-4} \cline{4-4} 
 & Literature & {\small{}Papers, textbooks, etc.} & {\small{}The Iron bar group used a damping term from a paper provided
by the instructor.}\tabularnewline
\hline 
\multirow{4}{0.75in}{\textbf{Products}} & Target model & {\small{}A model that is an outcome of a scientific inquiry} & {\small{}Describing the position of the iron bar oscillator as a function
of time.}\tabularnewline
\cline{2-4} \cline{3-4} \cline{4-4} 
 & Programs & {\small{}Code expressing intent to the computer} & {\small{}All groups turn in a Jupyter notebook that is executable
by the course instructor.}\tabularnewline
\cline{2-4} \cline{3-4} \cline{4-4} 
 & Documentation & {\small{}Comments, notes, and ancillary text facilitating use and
understanding of programs} & {\small{}Jupyter notebooks incorporated documentation integrated with
the code.}\tabularnewline
\cline{2-4} \cline{3-4} \cline{4-4} 
 & Publications & {\small{}Documents for disseminating results} & {\small{}All student groups produced reports on their findings and
presented their findings to the class.}\tabularnewline
\hline 
\end{tabular}

\caption{\label{tab:knowledge}\textbf{Examples of abstractions and artifacts} in the metamodel identified from student work}

\end{table*}

\begin{table*}
\begin{tabular}{|>{\centering}p{0.75in}|>{\raggedright}p{1in}|>{\raggedright}p{1.6in}|>{\raggedright}p{3in}|}
\hline 
 & \textbf{Practice} & \textbf{Definition} & \textbf{Example}\tabularnewline
\hline 
\multirow{6}{0.75in}{\textbf{Production}} & {\small{}Planning} & {\small{}Structure of code and algorithms are identified ahead of
implementation} & {\small{}The Iron bar group created a Python class prototype for organizing
their code.}\tabularnewline
\cline{2-4} \cline{3-4} \cline{4-4} 
 & {\small{}Implementation} & {\small{}Writing a program} & {\small{}All groups collaboratively wrote and implemented Jupyter
notebooks. }\tabularnewline
\cline{2-4} \cline{3-4} \cline{4-4} 
 & {\small{}Running} & {\small{}Executing a program to produce results} & {\small{}All groups executed programs to generate the results they
presented.}\tabularnewline
\cline{2-4} \cline{3-4} \cline{4-4} 
 & {\small{}Representing data} & {\small{}Processing data into a form amenable for computation or visualization} & {\small{}The Torsion group processed their data to be over a different
interval than the $-\pi$ to $\pi$ interval output by their sensor.}\tabularnewline
\cline{2-4} \cline{3-4} \cline{4-4} 
 & {\small{}Interpreting} & {\small{}Relating the outputs of a model to observable quantities
or phenomena} & {\small{}The Half-pipe group attributes the greater damping of the
heavier mass to greater movement of the ramp (and therefore energy
lost to that movement).}\tabularnewline
\cline{2-4} \cline{3-4} \cline{4-4} 
 & {\small{}Concluding} & {\small{}Claims about the outcome of the project based on production
and critique} & {\small{}The Iron bar group concluded that the effective potential
satiasfactorily modeled the behavior of their system.}\tabularnewline
\hline 
\multirow{6}{0.75in}{\textbf{Critique}} & {\small{}Testing} & {\small{}Comparing the output of the program to expected results or
data} & {\small{}Students in the Half-pope group continuously tested their
code in several phases as they modeled the ramp and then the motion
of the cylinder.}\tabularnewline
\cline{2-4} \cline{3-4} \cline{4-4} 
 & {\small{}Debugging} & {\small{}Editing the program to ensure it produces intended results } & {\small{}The Iron Bar group recognized that the BoxPlot tool cannot
represent their data in the way they intended.}\tabularnewline
\cline{2-4} \cline{3-4} \cline{4-4} 
 & {\small{}Numerical Analysis} & {\small{}Predicting or analyzing the performance of a program in terms
of cost, error or stability} & {\small{}The torsion group compares how the error between the numerical
and analytical solution change over time. }\tabularnewline
\cline{2-4} \cline{3-4} \cline{4-4} 
 & {\small{}Visualization} & {\small{}Conversion of data into a graphical display} & {\small{}The Torsion group produced multiple visualizations to verify
that their algorithm reconstructed the trajectory properly. }\tabularnewline
\cline{2-4} \cline{3-4} \cline{4-4} 
 & {\small{}Validation} & {\small{}Performing two or more different calculations and comparing
the results} & {\small{}The Torsion group compared both their numerical and analytical
models to their experimental data.}\tabularnewline
\cline{2-4} \cline{3-4} \cline{4-4} 
 & {\small{}Robustness Studies} & {\small{}Establishing the domain of validity of a computational model} & {\small{}The Half-pipe group used their computational model to analyze
the motion of different objects on their ramp.}\tabularnewline
\hline 
\end{tabular}

\caption{\label{tab:activities}\textbf{Examples of practices} in the metamodel identified from student work}

\end{table*}

This observation offers a strong motivation for encouraging student agency in class projects, whereby students are empowered to make decisions about what they want to think about and the ways they explore these curiosities \citep{engestrom2010studies}. In order for students to engage in many of the various activities that comprise computational modeling, as well as navigating the relations between them, it is also essential that the educational environment facilitates and promotes student \emph{agency}. Agency has attracted significant attention in science education and PER specifically, and will be discussed further in the Course Design section below. For now, we simply note that each connection between critique and production offers a potential opportunity for student agency, in addition to the choice of resources, objectives and products discussed earlier. We therefore direct our use of the metamodel toward situations that both expand and deepen the intellectual possibilities available to students to do science. 

One of our goals in this work is to design course projects that enable students to explore the totality of the computational modeling process, and so our design must aim quite broadly. If it is successful, we might expect to see in their work at least:

\begin{enumerate}
\item An identifiable target, i.e. an observation, a conceived or reported phenomenon, a known physical effect. 

\item Objectives for the computational work. This might include qualitatively describing the target phenomenon, interpolating between asymptotic cases, fitting data, producing a desired physical effect.

\item Standards that determine whether these objectives are met, such as the quality of fit expected, or which features of the phenomenon the computational work should display. 

\item Identification of resources from which the model is drawn. This includes general background knowledge, or the computational template, (e.g. Newton's equations) but also contextual knowledge required to apply the general theory or model to the target phenomenon and meet the objectives. 

\item Identification of ancillary phenomena that must also be modeled to meet the objectives. 
\end{enumerate}

This is not an exhaustive list, but some factors we shall use in our later analysis. We will later examine a selection of student work in Section \ref{sec:Student-Work} with reference to these criteria, and suggest further refinements in Section \ref{sec:Discussion}. 

\section{Course Design\label{sec:Design}}

Our epistemological account of computational modeling and the associated metamodel were used to inform the design of computational physics projects in the classroom. Building from a design research approach \citep{brown1992design}, we have theorized and designed a learning environment to support computational approaches to disciplinary learning in science, while also studying how the theoretical proposals unfold in activity. The broader project involves work from K-12, examining aspects of teacher learning, student learning, and the ways relationships to disciplinary 
practice, tools, and co-learners are renegotiated through making\citep{gravel2021}, as will be further discussed in section \ref{sec:making} below.

The design enactment described here is an extension and elaboration of the computational making approach in K-12 settings to the university computational physics course. This was driven largely by a conjecture that computational making would allow further investigation of how students move from observable physical behaviors in the world to computational models and solutions. By making space for students to act with agency both over the physical systems they build and over the computational modeling of those systems, we aimed to create an environment that mimics the professional practice of computational modeling in physics. 

\subsection{Epistemic and pragmatic agency}

As we discussed in the preceding section, a crucial requirement to provide students with the opportunity to perform the full range of activities we envisage as important to computational modeling is \emph{agency}\citep{engestrom2010studies}. Agency plays a key role in scientific practice because progress in science is not linear and there are often fundamentally different ways to solve problems, conduct inquiries in science, or conceive phenomena. Which models are protected and which ones are revisionable, for example, is based on practical working principles rather than on epistemological virtues as discussed in the previous section.

The philosopher Sophie Ritson has scrutinized the epistemic role that the interrelated idea of \emph{creativity} has on experimental physics, defining creativity as: ``the capacity to increase the epistemic value of a measurement by transforming the model of a routine measurement process''\citep{ritson2021creativity}. Agency offers to scientists the possibility of revising, iterating and improving the outcome of an inquiry, as well as its objectives and hence its ultimate epistemic value.  Agency therefore allows for creativity in experimental science in Ritson's sense. For pragmatists, inquiries are a means to meeting objectives, and hence there is a considerable role for creativity in doing science. The same argument there applies to computational modeling as considered here.

We also distinguish between students' agency over their pragmatic or practical goals (e.g. what materials should be chosen for constructing an oscillator) and agency over their epistemic goals (e.g. what qualifies as a sufficient agreement between the computational model and physical system). We refer to the former as \emph{pragmatic agency} \cite{hitlin2007time} and the latter as \emph{epistemic agency} \citep{miller2018addressing,stroupe2014examining}. These two constructs are intertwined and overlapping, as seen in our metamodel: students may make practical decisions about how best to plan and execute their code that have consequences for how they draw conclusions based on that code. Together, students' opportunities to make decisions and evaluate the outcomes of those decisions constitute agency \citep{alsop2014activism}, and contribute to conditions supportive 
of epistemic practices that sustain computational modeling activity. 

Within physics education research, the role of agency, including epistemic agency, has been studied in the context of undergraduate laboratories. \cite{holmes2020developing,smith2020expectations,kozminski2014aapt,phillips2021not,dounas2017student, dounas2018characterizing, etkina2010design}. While offering encouraging insights into how lab settings contribute to students' developing disciplinary practices, this work has largely not addressed the role of computational thinking or modeling. It is our contention that incorporating making can bridge deliberate efforts to center student agency with the goal of cultivating computational modeling activities.

\subsection{Computational Making}
\label{sec:making}


Our enlarged conception of computational modeling includes how scientists may go back and forth between seeing phenomena in the world to modeling those phenomena computationally. This suggests a focused and elaborated attention to the physical and material aspects of phenomena---to the ``stuff'' in the world that people engage with, have histories with, and can manipulate and explore. Making transforms these histories with materials into resources to be used in the computational inquiry. To operationalize these practices in our design, we briefly introduce the idea of \emph{computational making}, one of us (B.E. Gravel) has studied in K-12 settings \citep{gravel2022} . 

In our computational making environment, students are explicitly positioned as producing knowledge in the classroom through the selection and iterative manipulation of materials in conversation with computational tools, practices, and artifacts.  
Making is transdisciplinary and multimodal\citep{tucker2019stem}: students can leverage resources, knowledge and practices beyond just their prior experiences as physicists. Its generative and flexible nature opens space to examine the utility and explanatory power of the metamodel in designing improved environments for physics education.

Computational making specifically draws on attempts to define ``computational thinking''\citep{Wing.2006,grover2013computational,li2020computational} alongside elements of design and making \citep{wardrip2015learning}, to propose a kind of structured and iterative approach to learning in the processes of making\citep{knight2015making}. It reflects the nonlinear maneuvering between production and critique observed in scientific collaboration and discourse\citep{gravel2017integrating} and provides a multiplicity of paths for student work, all of which is aligned with our pragmatic construction of computational modeling in Section \ref{sec:EpistemicModel}.  In this way, the computational making environment provides opportunities for both epistemic and pragmatic agency, as they iteratively make and model and engage in inquiry and scientific practices. 

Prior examinations of computational systems pushing beyond programming in science learning justify our proposed approach. The production and critique of computational artifacts supports learning in disciplinary pursuits \citep{gravel2017integrating} expanding depictions of computation from programming and simulation-based representational environments into physical objects and arrangements of materials in the world\citep{blikstein2014bifocal, disessa1986boxer}. We note connections with constructionist ideas \citep{Papert2000} of learners as agents. Seen through a constructionist lens\citep{hoyles2002rethinking}, making provides an environment of ``powerful ideas''\cite{papert2020mindstorms} that affords new ways of thinking, of putting knowledge to use, and of making personal and epistemological connections with other domains of knowledge. We also leverage the idea of public artifacts\citep{harel1991constructionism} as an organizing principle, both as the products of making and the resulting computational model itself. A subsequent paper will examine the Making component of our design in more detail and we discuss connections between our work and other theoretical approaches to computation below in section \ref{sec:Comparison}. 

\begin{figure}
\includegraphics[width=1\columnwidth]{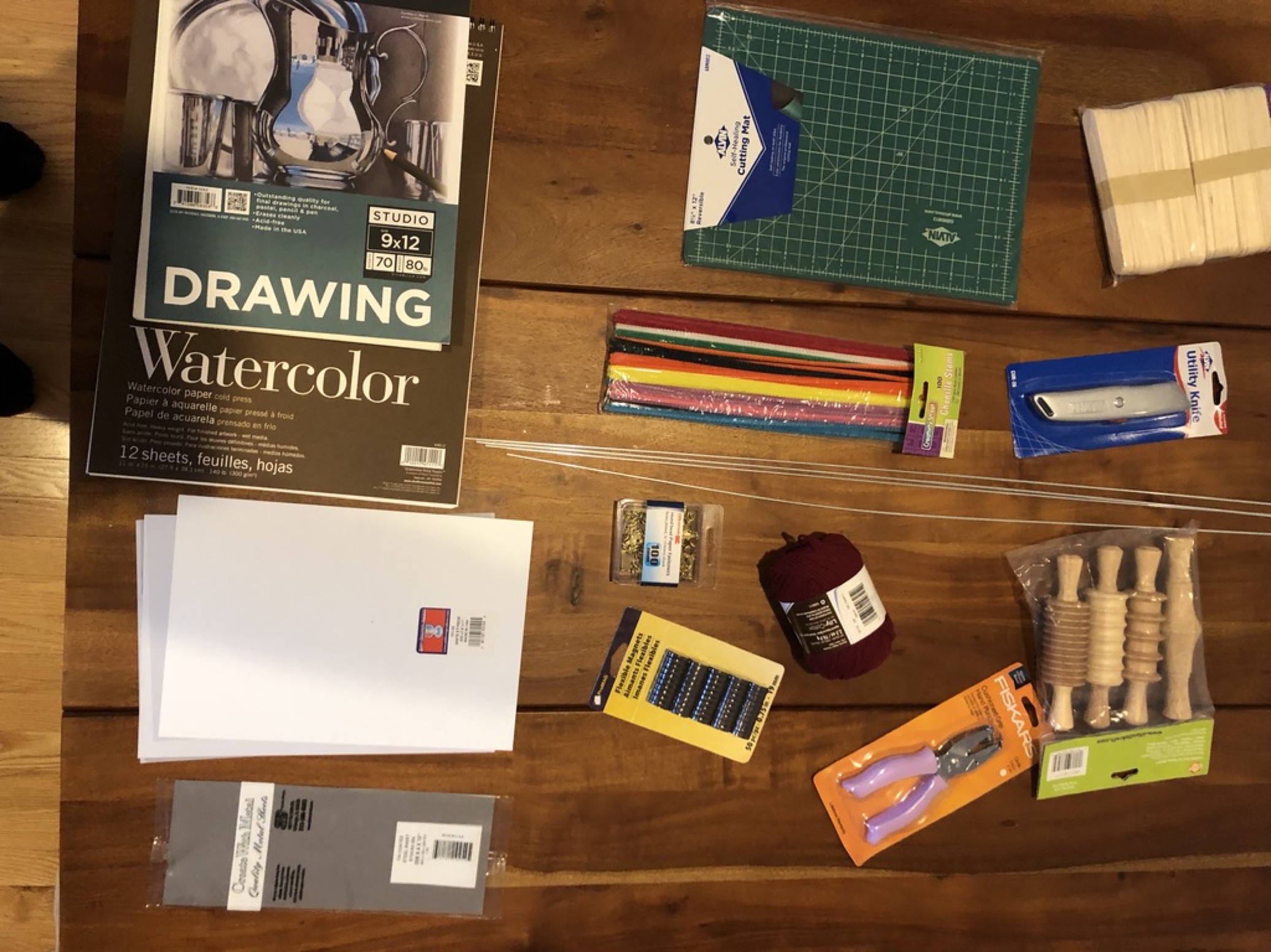}

\caption{\textbf{\label{fig:Materials}A selection of making materials} provided to the students. Materials are selected to provide a wide variety of properties, each be usable in more than one way and to avoid obviously resembling lab items such as springs, masses.}
\end{figure}

\subsubsection{Making Projects}

A making project in our Computational Physics course includes the following elements: 
\begin{description}
\item [{An~open-ended~prompt}] that invites exploration rather than focusing on a particular scenario. We have used\emph{ ``make something that moves''} in K-12 work and \emph{``make an oscillator''} as described below. The prompt could also incorporate a physical demonstration. 
\item [{Construction~of~a~physical~artifact}] from provided and found materials. Adequate time must be provided to explore materials, produce first iterations (or "drafts"), and revise the artifacts, including space to explore curious and unexpected material behaviors.
\item [{Quantitatively~characterizing~phenomena}] This could include image or video capture, cameras and tracking software e.g. \emph{Tracker}\citep{brown2007combining,brown2009innovative}, but might also involve logging and processing the output of other available sensors such as temperature or magnetic field\citep{staacks2018advanced}. 
\item [{Producing~a~computational~model}] of the particular artifact
constructed, incorporating decisions about representations, algorithms,
what to include, etc. 
\item [{Critique~activities~and~iteration}] including comparing simulated and experimentally
obtained data, revising, iterating and choosing alternative representations, and revisiting and revising the physical artifacts themselves. 
\end{description}

Making projects in other environments have enabled low-stakes, creative, playful exploration of phenomena, provide a rich opportunity to explore materials and material relationships, and flatten hierarchies such as those that exist between teachers and students\citep{vossoughi2020beyond}.  

The prompts are carefully chosen: Students confronted with a table of materials and a simple directive---\emph{Make an oscillator}---are positioned as designers, with agency over how to initiate the construction of an object, and the formulation of a problem \cite{phillips2017problematizing}. Students are thus offered a large range of possibilities for resources, activities, and even objectives to be considered.

As well as the initial prompt, a second design lever available to the instructor is the choice of materials provided. In selecting the materials, mostly obtained from a craft store, we specifically chose items that had a wide range of physical properties in terms of rigidity, malleability, texture. We also avoided objects that explicitly resembled ``masses'' or ``springs'' that might obviously recall textbook representations of oscillators. 

The projects take place over 4-5 class periods in a semester-long Computational Physics course offered to a mix of undergraduate and graduate students including Physics, Math and CS majors, all with some programming experience. The majority of class time is used for group work and the students work in teams of 3-4 students. 

In this paper, we focus on the third project in the course, which begins with a challenge \emph{``Make an oscillator''}. Students use a variety of provided materials to make as many kind of oscillators as they can, and present them to the class in the first session of the project. In subsequent sessions, the students refine, combine, or make completely new oscillators, characterize their motion experimentally, and build a computational model of one of their oscillators. They compare and refine both their experiment, taking additional data where necessary and improving imaging or changing the oscillator to test predictions, and the model, incorporating additional effects or comparing different aspects of the model with data. The activity was designed to provide a sufficiently open-ended task to enable students to identify phenomena of interest, articulate epistemic objectives, and have the opportunity to make revisions both to the model and the physical oscillator. It also incorporated numerous opportunities for students to share progress with the class and exchange ideas. 

\section{Implementation\label{sec:Implementation}}

\begin{table*}
\begin{tabular}{|c|>{\raggedright}p{2.5in}|>{\raggedright}p{2.5in}|}
\hline 
Session & Activities & Prompts\tabularnewline
\hline 
1 & Project introduction & Challenge to ``Make an oscillator'' from  \tabularnewline
 & Students make prototype oscillators. & instructor. \tabularnewline
 & Students share a prototype to the class. & \tabularnewline
\hline 
2 & Students familiar with Tracker give tips. &  ``Let's get some data''.\tabularnewline
 & Groups form and begin refining oscillators. &  Students invited to mix groups.\tabularnewline
\hline 
3 & Q\&A. Microlecture on documenting process. & Explanation of the importance of iteration \tabularnewline
 & Students work on refining oscillators and capturing data. & with reference to engineering practices. \tabularnewline
\hline 
4 & Group presentations on their progress. & Parameters for one slide presentation.\tabularnewline
\hline 
5 & Optional microlecture on Lagrangian mechanics. & Make a computer model\tabularnewline
 & Groups begin modeling. & ``Work out what equations we want to solve''\tabularnewline
\hline 
6 & Modeling and further refinement/data collection. & \tabularnewline
\hline 
7 & Class reflection on activity. & \tabularnewline
\hline 
\end{tabular}

\caption{\label{tab:MakeAnOscillator}\textbf{Class sequence} of `Make an Oscillator'
activity from Spring 2019 session, together with a summary of instructor prompts.}
\end{table*}

The Oscillator project described above was included in the Spring 2019, Spring 2020 and Spring 2021 offerings of the course; each of these iterations involved $18-30$ students. Slack was used as the primary communication platform for the course, with Python and Jupyter the recommended environment for computation. In-class activities consisted of supported group work, where the instructor would join each group 1-2 times per period, short ``microlectures'' on a key topic of interest and communication and reflection activities. In this paper, we focus on student projects from 2019, as that implementation of the course was most thoroughly documented through video recording and not disrupted by the COVID-19 pandemic. A timeline for each project reconstructed from observational data is displayed in \ref{tab:MakeAnOscillator}. The table briefly summarizes when verbal prompts were given to the students as reconstructed from the video data; further details are provided in the Appendix. For each project, student groups were asked to submit their code, a 1-2 page summary report and an individual self assessment\citep{Burke.2017}. Some groups also included videos and images of their oscillators as part of their submission. 

\subsection{Data Collection}

Observations of the class sessions involving computational making were conducted by members of the research team, and field notes\citep{emerson2011writing} were generated by each observer. Additionally, video recordings were made of students working around the materials table and in their project groups. Roving cameras were used\citep{leonard2013insight} to capture specific interactions such as students reconciling their computational model with empirical data through the re-interpretations of the physical oscillators they built. Finally, student presentations were recorded, and all work submitted for the course was collected for analysis.

\section{Analysis and Results}\label{sec:Student-Work}

We now operationalize the metamodel presented in Section \ref{sec:EpistemicModel} to analyze student work. Our objective in doing so is to better understand what students are doing in their projects, how they maneuver between production and critique practices and to identify possible improvements to the design. To achieve this, we present the results of a content analysis\citep{miles2014qualitative, vaismoradi2013content} of students' written work and code in the Oscillator Project introduced in Section~\ref{sec:Design}. Ongoing studies of the remaining data corpus continues, including deeper interaction analyses to map relationships between the activities structures, the metamodel, and students' discourse. Our process is based on a thorough review and accounting of the components students included in their assignments. Attention has been given to the particular physics principles students discussed (e.g. causes of energy dissipation), reference to particular aspects of the metamodel (e.g. resources, objectives, and products), computational approaches (e.g. Lagrangian methods) and descriptions of how they engaged in different computational modeling practices (e.g. iterations, validations, etc.). As our process is primarily descriptive and involves identifying features of student work that reflect items from our metamodel that we identified through philosophical constructs in addition to student and professional practice, our approach follows content analysis as described Vaismodari and colleagues \cite{vaismoradi2013content}.

Here we will trace the work of three groups, one that made a skateboard style half-pipe, one that made a pendulum with an iron bar and a magnetic bob, and one that made a torsion oscillator. Across these three groups, we first describe their pathway through the project. Then, we analyze their written work using the metamodel as a framework for understanding their activities and practices. Finally, we conclude with a summary of examples of other activities from our metamodel seen from these three groups. 

\subsection{Qualitative summary of groups' trajectories}\label{sec:qualitativesummary}

First, we offer a description of how three groups navigated the oscillator project. These descriptions are constructed from their written documents as well as video of the students working in class. Then, we present a thematic analysis of this work using the five main components of our metamodel--products, production, objectives, resources, and critique--as the organizing themes. 

\begin{figure*}
    \centering
    \includegraphics[width=1\textwidth]{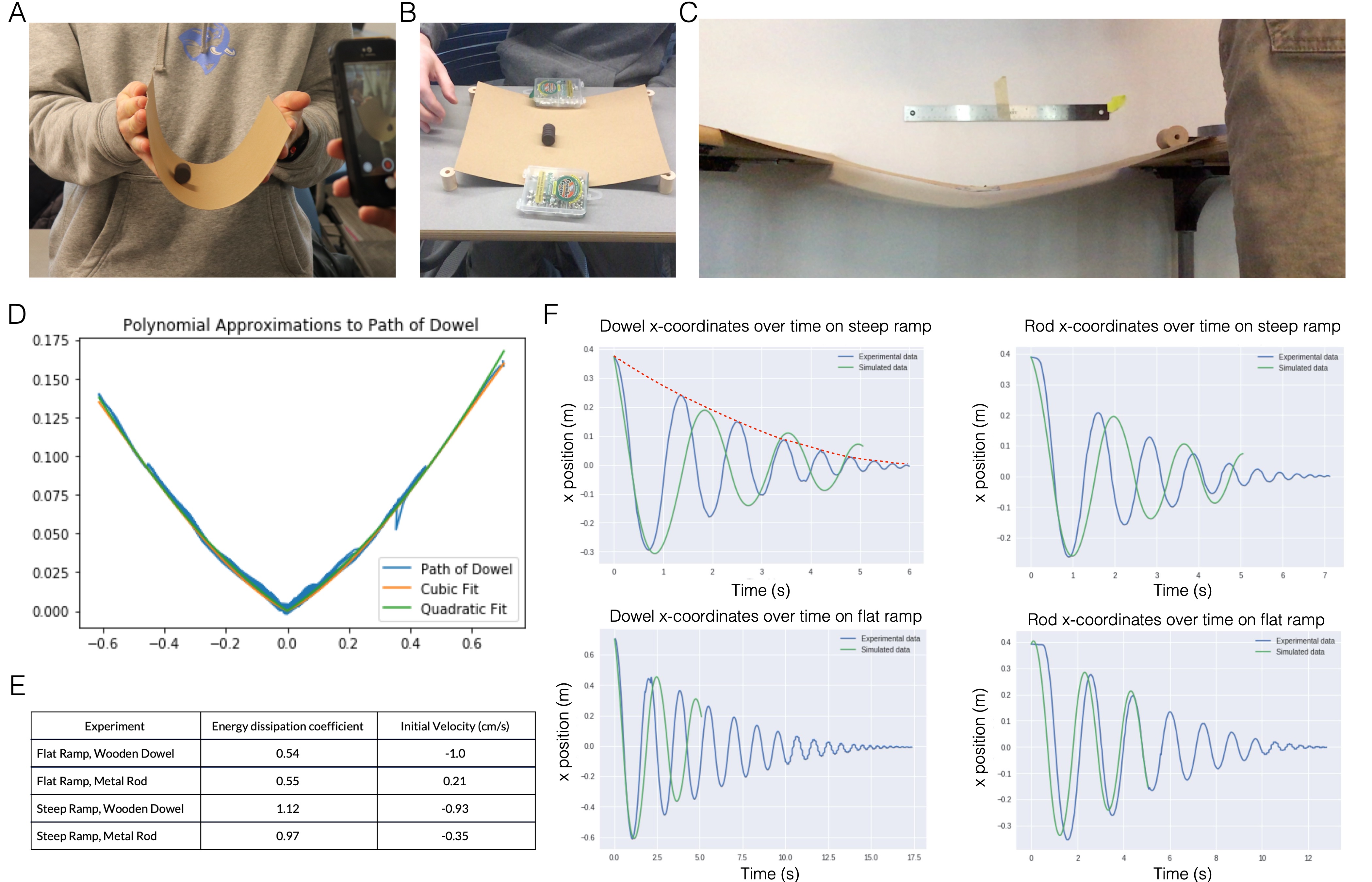}
    \caption{\textbf{\label{fig:halfpipe}Half-pipe group.} \textbf{A} Initial idea, \textbf{B} Second iteration, \textbf{C} Final production iteration of the oscillator. \textbf{D} Ramp profile fit to polynomial functions [Notebook]. \textbf{E} Table of extracted energy dissipation coefficients [Report]. \textbf{F} Fits of model to experimental data for different rollers and ramps[Report]. \emph{Note that text size has been increased for legibility purposes, however subfigure captions are verbatim what the students wrote. The red dashed line has been added to illustrate our interpretation of what the group refers to as an ``envelope''.}}
\end{figure*}

\begin{figure*}
    \centering
    \includegraphics[width=0.8\textwidth]{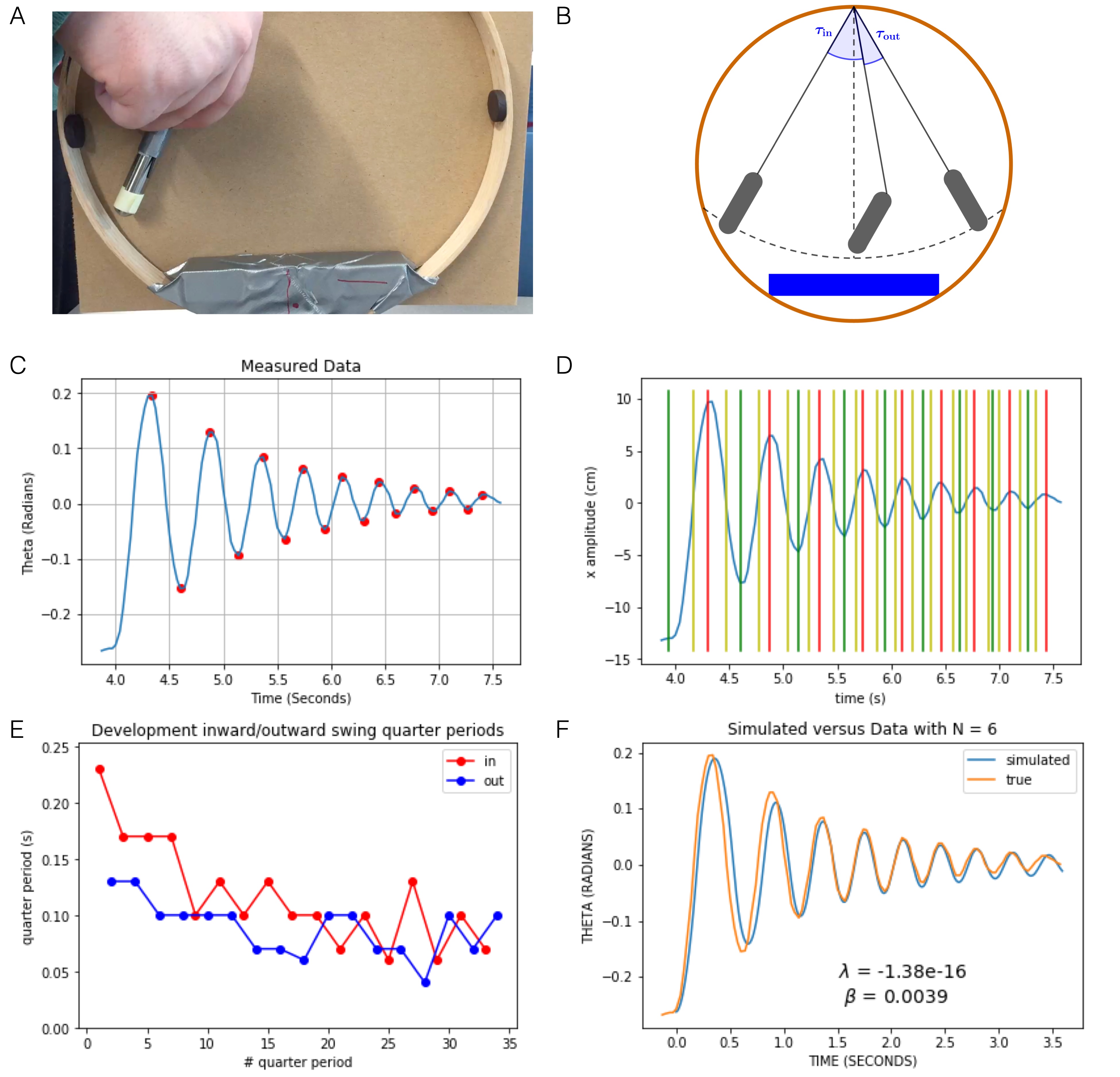}
    \caption{\textbf{\label{fig:ironbar}Iron bar group.} \textbf{A} Snapshot of oscillator; the bob is a magnet and the iron bar is held onto a wooden frame with tape. \textbf{B} Student-generated cartoon depicting oscillator configuration at different points in the motion; note tilted magnetic bob at base [Report].
    \textbf{C} Visualization showing maxima and minima detected by their algorithm [Notebook] 
    \textbf{D} Division of time into \emph{``quarter periods''} [Notebook] 
    \textbf{E} Length of quarter periods as a function of time for motion towards and away from the iron bar (\emph{``in''} and \emph{``out''}) [Report]
    \textbf{F} Final fitted data [Report]. Note that text size has been increased for legibility purposes, however subfigure captions are verbatim what the students wrote.}.
\end{figure*}

\begin{figure*}
    \centering
    \includegraphics[width=1\textwidth]{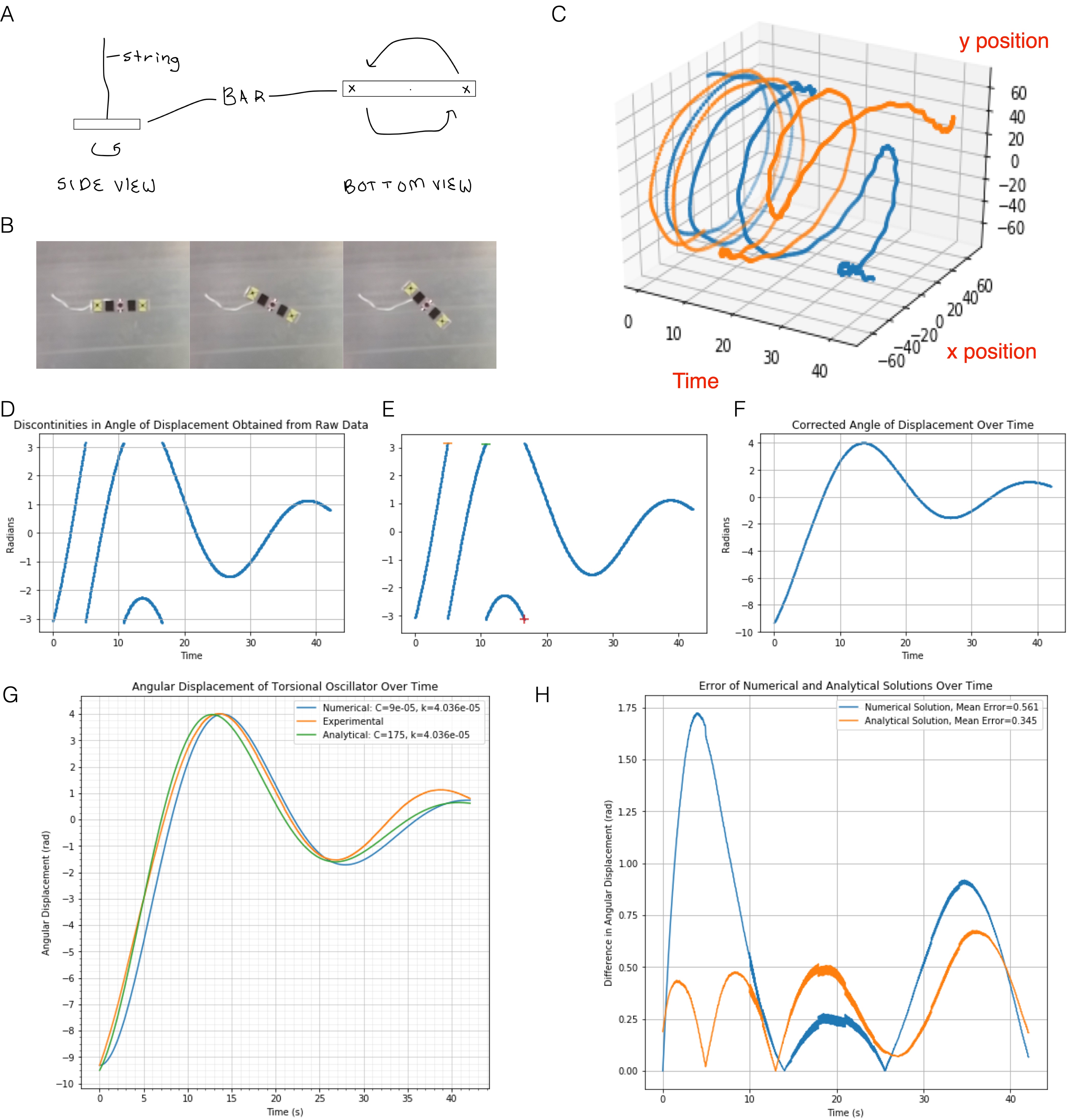}
    \caption{\textbf{\label{fig:torsion}Torsion group.} \textbf{A} Sketch of the torsional oscillator. \textbf{B} Snapshots from a submitted movie of the oscillator rotating. \textbf{C} Visualization of the position of two tracked points as a function of time. \emph{Note text in red was added here to aid clarity}. \textbf{D} Angular displacement $\theta(t)$ as initially reconstructed. \textbf{E} Intermediate plot displaying where their algorithm identified discontinuities. \textbf{F} Final continuous angular displacement function. \textbf{G} Numerical, analytical and experimental data. \textbf{H} Plots of the discrepancies |numerical-experimental| and |analytical-experimental|. }
\end{figure*}

\subsubsection{The Half-pipe Group}\label{sec:halfpipe}

The Half-pipe group decided to base their oscillator on a skateboard ramp, examining how a cylindrical object would move after being released on one side. They began their design with a simple piece of cardstock, held between two hands (Fig \ref{fig:halfpipe} A). They quickly recognized that they needed to fix the edges of the paper, and decided to elevate the two sides of the cardstock on a table (Fig \ref{fig:halfpipe} B). They collected video data in this configuration with both a wooden dowel and a metal rod as the cylinders. They then explored a steeper ramp (Fig \ref{fig:halfpipe} C) with the same two objects. This steeper ramp was made up of two pieces of cardstock taped together at the bottom, creating a slight `V' shape. They took data for the four possible combinations of metal vs wood cylinders and high vs low slope they created, and compared these data to one another. 

The group used the Lagrangian framework to build their model, with Euler-Langrange equations, 

\begin{equation}
    \frac{d u}{d t} = - g \frac{d h}{d x} - \beta u, \frac{d x}{d t} = u, \label{elhalfpipe}
\end{equation}
where $x$ is the horizontal position of the roller and $u$ its time derivative. The function $h(x)$  describes the height of the ramp as a function of $x$ and $\beta$ is an energy dissipation coefficient that combines multiple possible sources of dissipation. They fitted $h(x)$ using a polynomial from experimental data to a piecewise polynomial and reported this fitting process separately (Fig. \ref{fig:halfpipe}D). Having found a suitable form for $h(x)$, they integrated Eq. (\ref{elhalfpipe}) using a forward Euler integrator from initial position $x_0$ and velocity $u_0$ and created multiple visualizations, including animations, of the motion.

As they began to compare the predictions of their computational model with their experimental data, the Half pipe group found they could not bring the two into good visual alignment by adjusting the free parameters $x_0$, ${u_0}$ and $\beta$. They speculated that this was because of physical effects not included in their model. Instead, they decided to focus on determining the drag dissipation coefficient $\beta$ from the \emph{``envelope''} of the trajectory, i.e. the spatial positions of the stationary points (see dashed red line that has been added as an annotation to Fig. \ref{fig:halfpipe}F, upper left panel). They adjusted the parameters of their model to visually align the envelopes of the predicted and experimental trajectories and reported the fitted values of $\beta$ in a table (Fig. \ref{fig:halfpipe}E) together with the final plots (Fig. \ref{fig:halfpipe}F). The group concluded their report by articulating future objectives they would have wished to investigate if they had more time. 

\subsubsection{The Iron Bar group}\label{sec:ironbar}

The Iron Bar group began exploring how to create an oscillator out of magnets and magnetic materials. As displayed in Fig. \ref{fig:ironbar}A they settled on a design involving a bar magnet arranged vertically as a bob pendulum swinging over an iron bar arranged horizontally. They were able to visually observe the damping effects of the magnetic interaction as well as a difference in the behavior of the magnet on the down swings and upswings of the oscillation (see Fig. \ref{fig:ironbar} B). To understand the differences between the upswing and downswings of the oscillator, they broke the motion into \emph{``quarter periods''} to analyze separately (see Fig. \ref{fig:ironbar} C-E). 
To model this situation, they used the Lagrangian framework and attempted to build a computational model incorporating several physical effects. To construct the Euler-Lagrange equations, they used an analytical expression for the moment of inertia to compute the kinetic energy and included a gravitational potential, 
\begin{equation}
    V = m g R (1-\cos\theta).
\end{equation}

They also incorporated damping using a Rayleigh dissipation function from a paper provided by the instructor\cite{weigel2016predicting}, 
\begin{equation}
    D = \beta \dot{\theta}^2,
\end{equation}
where $\beta$ is a damping coefficient that could account for multiple damping forces. They solved the resulting E-L equations using a backwards Euler method. They first used this computational model to estimate the damping coefficient $\beta$ by visually fitting the output of the program to the data, but could not achieve a satisfactory fit. After this, they included an \emph{effective} potential into their E-L equation, 
\begin{equation}
    \frac{\lambda}{(R(1-\cos\theta)+R_0)^N}\label{ansatz}
\end{equation}
to account for the influence of the magnet. They estimated the parameter $R_0$, representing the closest point of approach that the magnetic bob makes with the bar, from the geometry of the system, and treated $\lambda$ and $N$ as fitting parameters. They were able to achieve a visual fit that they considered satisfactory with this additional effective potential term (Fig. \ref{fig:ironbar}F). 

They noted that their ability to check their model against their data was limited by the frame rate of their video data. Their analytical approaches did not account for this, so their final write up includes the binned experimental data.

\subsubsection{The Torsion oscillator group}\label{sec:torsion}

The Torsion group began experimenting with different objects hung from string. They tried several spinning objects before settling on a horizontal metal bar, which was easier to capture with video and Tracker (Fig. \ref{fig:torsion} A-C). As tracker initially output data from $-\pi$ to $\pi$, they had to process their data before modeling them in order to get a continuous trajectory from the discontinuous distribution they first obtained as \emph{``raw data''}, as can be seen on Fig. \ref{fig:torsion} D-F. 

Like the Iron Bar group, the Torsion group conducted both analytical and numerical analysis. They modeled their system as a torsional harmonic oscillator: they used the Newtonian framework and Hooke's law to construct a differential equation for the oscillator's motion, 
\begin{equation}
    I \frac{d^2 \theta}{d t^2} + C \frac{d \theta}{d t} + k \theta = \tau(t), \label{torsioneq}
\end{equation}
where $\theta$ is the angular position, $I$ is the moment of inertia, $C$ is an angular damping constant, $k$ is a spring constant to be fitted and $\tau$ is a driving torque. They used an analytical result for $I=\frac{1}{12} m (l^2+w^2)$ using the measured length $l$ and width $w$ of their bar and treated $C$ and $k$ as parameters to be estimated from experimental data. They discussed how the model Eq. (\ref{torsioneq}) may not accurately predict the motion of their oscillator, which was attached to regular craft string made up of multiple fibers. 

The Torsion group solved Eq. (\ref{torsioneq}) using the Runge-Kutta integrator in \texttt{SciPy}. They also considered an analytical solution, 
\begin{equation}
    \theta = A e^{-\alpha t} \cos(\omega t+\phi),
\end{equation}
where $A$ the amplitude, $\omega$ the damped resonant frequency, $\alpha$ the reciprocal of the damping coefficient and $\phi$ the phase shift are all parameters to be fitted.

They then compared the quality of the analytical and numerical fits to data, as presented in Fig. \ref{fig:torsion}G-H. They observed, from their examination of the fits, that both methods failed to model the oscillator after the first few seconds. Note that in their work, the students mislabelled ``frames'' as ``seconds.'' This discrepancy may account for very different fit parameters they found in analytical and numerical models. 

\subsection{Content Analysis}

A key aspect of our metamodel is that the practices involved in critique and production are closely tied to the resources, products, and objectives deployed. These items are therefore best understood through their connections to each other. For this reason, we organize our analysis of student work focusing in on three items within the metamodel, while describing their connections to other items in our analysis: 1) the \textbf{critique} practice of visualization, 2) the \textbf{resource} of standards of evidence, and 3) epistemic \textbf{objectives}. We use these as three organizing themes for thematic content analysis. Within the analysis below, we \textbf{boldface} items from the metamodel to highlight the interconnections.

This analysis has a dual purpose: we argue for the validity of our metamodel given we see the items from the model in student work, and we show how we can use our metamodel to understand aspects of student work. We could achieve the same aims using different items from the metamodel: it would be possible to organize the analysis around \textbf{data} as a theme and analyze how students \textbf{represent} and \textbf{visualize} that data. We have chosen the three themes of visualization, standards of evidence, and epistemic objectives primarily to minimize repeated analysis of the same excerpts of student work. We also have clear evidence of these themes in student reports and Jupyter notebooks. Some practices, such as debugging, are not always discussed in these products as students delete or edit non-functioning code. Similarly, the students in this iteration of the course did not consistently cite literature they consulted during their project, nor did they consistently describe their planning processes. 

\subsubsection{Visualization}

\begin{figure}
\includegraphics[width=1\columnwidth]{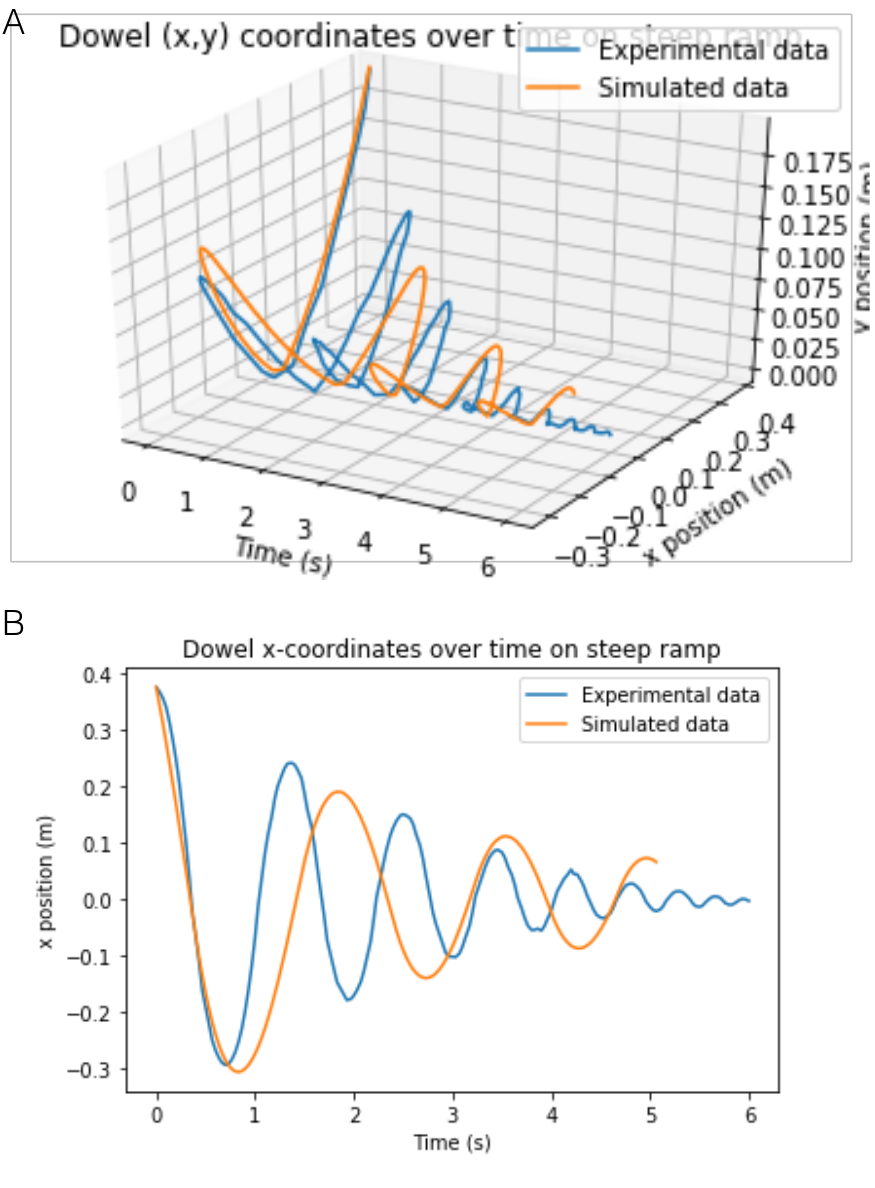}

\caption{\textbf{\label{fig:Visualizationhalfpipe}Students' intermediate visualizations} of experimental and simulated trajectories from the Half-pipe group's report. Legend position on 7A is from the student work. The text behind the semi-transparent legend reads ``Dowel (x,y) coordinates over time on steep ramp."}
\end{figure}

All student groups produced multiple \textbf{visualizations} during their work following \textbf{production} activities like gathering data or running the program. Visualizations allowed them to \textbf{critique} their data, what their code was computing and the performance of their code. The visualization process often shaped the \textbf{epistemic} and \textbf{pragmatic objectives} of the inquiry. 

As described earlier, the Torsion group used multiple visualizations in the course of reconstructing a continuous experimental trajectory. They examined raw data from Tracker with a visual representation (Fig. \ref{fig:torsion}E) and then showed a different representation (Fig. \ref{fig:torsion}D) after their initial processing. They used a visual representation to \textbf{validate/cross check} that their algorithm had found discontinuities in $\theta(t)$ correctly (Fig. \ref{fig:torsion}E). Their final assessment that the embedded modeling process had succeeded was based on a plot of the final reconstructed trajectory (Fig. \ref{fig:torsion}F).

Visualization played a key role in the Half-pipe group's decision to change the \textbf{epistemic objective} of their project. They describe how they produced multiple visual representations in their notebook: 

\begin{quote}
\emph{``In the steps below, we analze [sic] each of our for [sic] experiments individually. We play around with the beta-term (energy-loss coefficient) and other constants (e.g. initial velocity) to fit our simulation to the observed data. We visualize the observed and simulated paths side-by-side in a variety of graphical manners.''} [Half-pipe group, Jupyter notebook]
\end{quote}

Examples of intermediate plots they produced in their notebook are shown in Fig. \ref{fig:Visualizationhalfpipe}. These visual artifacts were central to the group's assessment of whether their \textbf{epistemic objective} had been met and hence whether their inquiry had been successful. Their \textbf{standard} of success was whether there was satisfactory agreement in the visual superposition of their experimental data with their simulated trajectory. 

In this case, they determined that they could \emph{not} adequately model the actual motion and changed their \textbf{epistemic objective} as described above. With the scope of the inquiry restricted to matching the envelopes of the trajectories, the Half-pipe group used similar plots displayed in their report (Fig. \ref{fig:halfpipe}F) to provide evidence that their modeling had met this more limited objective; hence these visual representations became part of the \textbf{product} of the inquiry. 

The Iron Bar group noticed from their video and by observing their apparatus that the orientation of the magnet changed depending on whether it was moving away from or towards the iron bar. They incorporated this into a ``cartoon'' of the motion Fig. \ref{fig:ironbar}B. 

Using their \textbf{expertise} and \textbf{background knowledge} to hypothesize that the change in orientation affected the damping behavior, this group set a secondary \textbf{epistemic objective}:
\begin{quote}
\emph{``[To] see if quarter-periods in which the pendulum was swinging inwards towards the iron block were of a different length than quarter-periods when it was swinging away from the block.''}[Iron Bar group, Report]
\end{quote}
To \textbf{validate} this, the group first \textbf{produced} code that determined the minima and maxima of the trajectory and \textbf{tested} that this had worked with a visual representation (Fig. \ref{fig:ironbar}C). They next used the results to create a visual representation that emphasized the duration of the portions of the motion towards and away from the iron bar (Fig. \ref{fig:ironbar}D). Dividing the motion into \emph{``quarter periods''}, they plotted the duration of these in order in Fig. \ref{fig:ironbar}E. This allowed them to \textbf{conclude},
\begin{quote}
\emph{``Periods converge over time but initially there is a large difference between inward swinging and outward swinging quarter periods.''} [Iron Bar group, Report]
\end{quote}
thus achieving the \textbf{epistemic objective} of their subsidiary inquiry. 

With these examples, we see \textbf{visualizations} as important things in their own right. Within the context of a complete inquiry process---including in embedded inquiries---they often have their own distinct \textbf{objectives}. Further they arise both in a \textbf{critique} mode, e.g. as resources are used to validate the computational model, and also to communicate the outcomes of the inquiry as part of the \textbf{products}. 

\subsubsection{Standards of evidence}

As the groups engaged in \textbf{critique} activities, they used various \textbf{standards of evidence} as \textbf{resources} to determine whether the evidence they had acquired met their \textbf{objectives} or not. 

The Half-pipe group used a qualitative \textbf{standard} for agreement, whether model and data could be satisfactorily superposed visually, in modeling the shape of the ramp (Fig. \ref{fig:halfpipe}D:
\begin{quote}\emph{``Visually, the cubic approximation as seen above more closely aligns with the actual path taken by the dowel, especially at the far left and right end. Thus, the cubic approximation will be used for the flat ramp configuration.''} [Half-pipe group, Jupyter notebook]\end{quote}
As discussed earlier, they could not meet this same standard for assessing the success of the broader modeling process:
\begin{quote}\emph{``getting the simulated position vs function curve to align with the observed data was impossible''} [Half-Pipe group, report]\end{quote}
They instead stated a new qualitative \textbf{standard} for agreement between model and experiment with reference to their revised \textbf{objectives}:
\begin{quote}\emph{``if the position vs. time curves aren't superimposed upon one another, as long as the crests are reaching the same heights, we have satisfactorily modeled the scenario, even if the crests are not simultaneous with respect to time.} [Half-Pipe group, report]\end{quote}

Other groups used similar \textbf{standards of evidence}: the Iron bar group performed a manual fit by adjusting parameters and reported their fitted values in the report (see Fig. \ref{fig:ironbar}F). The Torsion group similarly displayed superposed simulated and experimental trajectories (Fig. \ref{fig:torsion}G) and additionally plotted residuals between models and data (Fig. \ref{fig:torsion}H).

The Iron bar also compared data with and without the iron bar to provide evidence that the motion of the magnet was more strongly damped when the iron bar was present:
\begin{quote}
\emph{``It does appear however that this dampening seems to be very prominent at maximum angular velocity, when the magnet is close to the iron bar. Second, the period of oscillation is distinctively shorter than the period of the same oscillator without the magnet. This shows that the magnetic interaction is indeed present and having a strong effect on the oscillator.''} [Iron Bar group, Jupyter notebook]
\end{quote}
They hence used both qualitative evidence, superposed trajectories in a \textbf{visualization}, and quantitative evidence, calculated periods, to justify including the effective potential in their model. We  see this exercise as an embedded inquiry in its own right, the results of which become a \textbf{resource} they used to justify aspects of the broader inquiry. 

The Half-pipe group articulated reasons why their model could not fully predict the observed trajectories by enumerating physical effects they had not incorporated into the model: 
\begin{quote}\emph{``We believe that this is due to the enumerable variables not accounted for by our simulator: air resistance, angular momentum, etc.} [Half-Pipe group, report]
\end{quote}
In doing so, they are implicitly identifying the disagreement between model and data as due to systematic effects, rather than random error. Rather that try to incorporate these missing components in the model, they chose to narrow their inquiry and re-frame their \textbf{epistemic objectives}. We suspect that this decision was shaped by pragmatic constraints such as the availability of resources (background knowledge, time) because the group identified missing effects as \emph{``Future objectives''} in their report. Specifically, they wanted to
\begin{quote}\emph{``
investigate alternative equations for energy loss (e.g. energy loss proportional to square of velocity, energy loss proportional to acceleration, etc.)''} [Half-Pipe group, report]
\end{quote}
and
\begin{quote}\emph{``
find a roller with a mass similar to that of the metal dowel but with a much larger moment of inertia; this would allow us to separate the effects of greater roller mass and greater roller moment of inertia''} [Half-Pipe group, report]
\end{quote}
We interpret their decision to re-center their inquiry on focusing on reproducing the envelope of the trajectory as one that could encapsulate many effects into a single energy loss coefficient that could be more simply measured and modeled. 

The Iron bar group, in contrast, made a conscious effort as they iterated between  \textbf{production} and \textbf{critique} to control systematic effects:
\begin{quote}
\emph{``A second iteration of this oscillator was constructed, replacing the string with two plastic cables to stop any radial or rotational motion of the pendulum.''} [Iron Bar group, report]
\end{quote}
as did the Torsion group:
\begin{quote}
\emph{``We began with a wooden ring, suspended using a rope, that would rotate around the center of the circle. However, in order to make data collection and physical modeling easier, it helped to have a heavier mass with more continuous surface to video-capture''}[Torsion group, report]
\end{quote}

In these examples, we see the groups using \textbf{standards of evidence} to assess whether their \textbf{objectives} have been met, to justify aspects of their model, and making a conscious effort to articulate or control for systematic effects as a result of \textbf{critique}. Nonetheless, as we discuss below, this aspect offers further opportunities for refinement of our instructional design. In future iterations of the design, we might provide students with additional \textbf{resources} to more clearly quantify their standards of evidence, to characterize and control systematic effects, and to provide uncertainty estimates as part of the \textbf{products} of their inquiry.

\subsubsection{Epistemic objectives}

Two of the three groups, the Iron Bar group and the Half-Pipe group, documented how their \textbf{epistemic objectives} shifted during their work in their reports and Jupyter notebooks in detail.

The Iron Bar group describe two initial \textbf{epistemic objectives} in the form of research questions: 

\begin{quote}\emph{``Can we find a reasonable model for the magnetic potential of the pendulum in order to accurately fit the data? Our second was a bit more abstract: Is the magnet doing something to this system that we can't model?''} [Iron Bar group, report]
\end{quote}

The first question is an elaboration of the assigned task, to model the behavior of the oscillator, and identifying the influence of the magnet as a particular target. To meet this objective, they incorporated the magnetic effect through an effective potential Eq. (\ref{ansatz}) in the Lagrangian framework.  They described that simplifications were needed to do so: 

\begin{quote}
\emph{``For simplification, assuming the iron bar to be infinite a guess at the potential introduced by the magnetic field could be of the form }[see Eq. (\ref{ansatz})]\emph{ The reasoning behind this potential form is that the strength of the magnetic field decreases with distance, and the $R_0$ term represents the minimum distance between the bob and the iron bar in the model we build.''} [Iron Bar group, report]
\end{quote}

This group did not possess the \textbf{resources} necessary to construct a more detailed model of the magnetic interaction, so for \emph{``simplification''} they chose to use their \textbf{expertise} to deduce an approximate form---a \emph{``guess''} as they put it---but nonetheless one consistent with their expectations for how a model of the magnetic interaction should behave. The acceptability of this sub-model ultimately was determined by its utility in meeting their \textbf{objectives} for the broader inquiry. The group found that doing so enabled them to successfully model the behavior of the magnetic pendulum and hence they considered that they succeeded in their first \textbf{epistemic objective.}

Their second \textbf{epistemic objective}, which they describe as more \emph{``abstract''}, was to identify features of their system that are not captured by their model. While they initially identified this as a target for understanding, their conclusions reflected a different stance towards this goal (here $N$ is a parameter from the effective potential):

\begin{quote}
\emph{``The equation for angular acceleration derived in this project does, at larger than expected values of $N$, simulate the physical system we created in class. Determining the exact reasoning behind this would be a fun project however due to time constraints could not fall within the scope of this assignment.''} [Iron Bar group, report]
\end{quote}

The Iron bar group describes how their value of $N$ is surprising but determine that understanding the detailed physical mechanism is beyond the scope of the assignment.  This group therefore abandoned their second \textbf{epistemic objective} of tying behavior of the magnet to specific physical phenomena.

This shift appears to emerge because they felt they had met their first \textbf{epistemic objective}: they were satisfied with the quality of fit between model and data. They also lacked an important \textbf{resource}, the time needed to carry out the additional work necessary, and hence made the decision that they did not need to go above and beyond the stated requirements of the assignment, satisfying the implied \textbf{pragmatic objective} of completing the work on time.


The Half-pipe group began with an initial \textbf{epistemic objective} of creating a computational model that exactly matched the data. However, they  shifted their \textbf{epistemic objectives} after failing to obtain a high quality fit between their data and computational model, as identified through \textbf{visualizations} (see Fig. \ref{fig:halfpipe}F):

\begin{quote}\emph{``We quickly learned that no matter how hard we tried, getting the simulated position vs function curve to align with the observed data was impossible. We believe that this is due to the enumerable variables not accounted for by our simulator: air resistance, angular momentum, etc.} [Half-Pipe group, report]
\end{quote}

They inspected their fits in order to \textbf{interpret} their results. They identified that the computational model did appear to capture the feature of energy dissipation. They decided to focus on matching the envelope of the model and experimental trajectories instead: 

\begin{quote}\emph{``However, in order to simulate the energy loss, we only cared that the total energy at any given point in time matched between the observed data and theoretical model. Thus, when assigning and tweaking the energy dissipation coefficients, we were more concerned with making the observed data and simulated model share common upper and lower envelopes. In other words, even if the position vs. time curves aren't superimposed upon one another, as long as the crests are reaching the same heights, we have satisfactorily modeled the scenario, even if the crests are not simultaneous with respect to time.''} [Half-Pipe group, report]
\end{quote}

The Half-pipe group determined that they could not meet their original \textbf{epistemic objective} according to their (implied) criterion for quality of fit. Rather than abandon the inquiry, this group realized they had the \textbf{resources}, including \textbf{expertise}, \textbf{background knowledge}, available \textbf{data}, etc., to model a sub-phenomenon of the oscillation, the energy dissipation coefficient (see Fig. \ref{fig:halfpipe}E, F). They therefore changed their \textbf{epistemic objective} to this new target. They were able to reuse the computational model already created, running their program repeatedly with different parameters to model the total energy loss.

These examples illustrate the fluidity of both scientific inquiries and models, which can evolve dynamically as the work proceeds: In both cases, initial iterations of \textbf{production} of models and data were followed by \textbf{critique}, including \textbf{visualization}, with reference to the groups' respective \textbf{epistemic objectives} as well as \textbf{pragmatic objectives} such as completing the inquiry in the time available. As a result of this, both groups felt that they lacked the \textbf{resources} to undertake the full program as originally envisioned. Instead, they altered the scope of their inquiry by dropping a secondary objective, in the case of the Iron Bar group, or by restricting the initial objective to something they believed they could achieve for the Half-pipe group. Their inquiry then proceeded with modified objectives, but deployed \textbf{resources} already available---including their code and data---to the new target.

\section{Discussion}\label{sec:Discussion}

Our metamodel is based upon philosophy of science \cite{humphreys2004extending, beauchemin2017autopsy,ritson2021uncertainty,ritson2021creativity, gelfert2016science}, research in PER on modeling and other metamodels \cite{Hestenes.1987, Brewe.2008,Zwickl.2014lgj,Zwickl.2015,Odden.2019,Rios.2019,Dounas-Frazer.2016,dounas2017student} and prior work on computational physics education, both by our group \cite{Burke.2017} and others \cite{Caballero.2015,Chabay.2008,Caballero.2018,Caballero.2019,caballero2012implementing}. Through content analysis of student work, we show that our metamodel accurately reflects computational physics practices we see in student work. 

In this analysis of student work, we are able to understand students' practice of computation within a learning environment where students were afforded agency both about what to model and how to model. The highly non-linear structure of our metamodel allows us to see the ways by which students make progress through their computational work, and in particular the maneuvers that they make. As students engage in various practices, they generate different types of artifacts and knowledge, and their epistemic and pragmatic goals evolve over time. 

The interplay of making physical objects and computational modeling, which we refer to as \emph{computational making} allows students to engage in an act of scientific inquiry in which they use the practices of physics: They work to collect data, analyze and simplify physical models and draw conclusions about how their physical apparatuses work. As scientists do, they use resources and produce products communicating their results and making them available for further inquiry. In this way, we see students engage in computation not only as a tool to use while doing physics but \emph{as a core part of what it means to do physics}.

While we have shown that the computational making environment is particularly well-suited to supporting students' in engaging in a wide range of computational physics practices (as discussed in more detail below in Sec \ref{Sec:ImpliInstr}), we expect to see some of the identified practices in other less open learning environments, consistent with the work of Caballero and colleagues \cite{Caballero.2015,Caballero.2018,Caballero.2019,caballero2012implementing}. We anticipate our metamodel also offers new ways to design instructional elements that target subsets of scientific inquiry. For example, an introductory course might scaffold assignments that focus on various aspects of inquiry, building up to a complete act of inquiry. Existing assignments might be redesigned to emphasize some of the practices identified and offer additional opportunities for agency. 

\subsection{Comparison with other Theoretical Approaches to Computation}
\label{sec:Comparison}

We now briefly consider our metamodel with reference to other theoretical approaches to computation that are being used in the Physics Education community and beyond. We stress that we do not believe that any single theoretical lens (including ours) is sufficient to describe a phenomenon as complex as computation, which by its nature is transdisciplinary, involves multiple representations and kinds of activity, and used for many different purposes even within physics as we noted in the Introduction. Rather, we see these theoretical lenses as both highlighting certain aspects of computation, and facilitating the organization of communities of researchers interested in advancing our understanding of computation. A more detailed comparison of these perspectives is left to future work.

The first of these, \emph{Constructionism}, has been extremely influential in Computing Education and has been exploited in part by other physics projects such as PheT\cite{perkins2006phet} and VPython\cite{scherer2000vpython}. Originally articulated by Seymour Papert\cite{papert2020mindstorms,Papert2000}, constructionism as a pedagogy involves students creating digital things such as code and documentation, as well as physical objects\cite{laurillard2020significance}. Constructionism has influenced the design of our environment rather than the metamodel as we shall discuss in more detail in a future paper. For example, computational making honors constructionist commitments to learners as agents, emphasizes the creation of shared things, and aims to include historically marginalized groups. However, constructionism by design embraces epistemological pluralism\citep{Turkle.1992}, i.e. it evades a specific formulation of knowledge production for the purpose of advancing equity, enabling students to bring their own knowledge to the classroom. We try to retain this aspect---a student's prior knowledge is a resource in our metamodel---while nonetheless providing a detailed metamodel of scientific inquiry.

A second lens, \emph{Computational Thinking}, emerged from an opinion piece by Jeanette Wing\cite{Wing.2006} who argued that many computer science practices, e.g. thinking recursively, abstraction and decomposition, constituted a ``universally applicable attitude and skill set'' that should be taught to others outside the discipline. It has become very popular in the STEM and Computer Science education literature in recent years\cite{li2020computational}, and there is an emerging body of work in Physics that has used these ideas productively\citep{orban2020computational, lyon2020computational}. Nonetheless, Computational Thinking has some important limitations: It was not formulated with reference to physics, and hence does not incorporate physics practices. Further, it does not provide a coherent description of how Computer Science practices inform one another to advance knowledge, even within Computer Science. That is to say, it is not an exclusively Computer Science equivalent to our metamodel. Rather, we believe that Computational Thinking represents a potentially rich source of practices\citep{weintrop2016defining} that could be incorporated into our metamodel in both production and critique modes. 

The final lens considered here, \emph{Computational Literacy}, is grounded in the work of Andrea DiSessa who theorized computation as a new type of literacy that fundamentally changes how students think and learn\cite{disessa2001changing,disessa2018computational}. The literacy metaphor emphasizes communication and social processes. Hence, Computational Literacy may call our attention to the kinds of resources available to the students, products they produce, as well as how students communicate in the process of an inquiry. Odden and coworkers have recently demonstrated DiSessa's idea of ``computational essays'', notebooks that combine code, text and figures to communicate an idea, as tools to scaffold professional physics practice\cite{Odden.2019,Odden.2020993}. Seeing these as documentary evidence of an inquiry, our metamodel provides additional tools to interpret epistemic maneuvers made by the students mirroring our content analysis here. Equally, Computational Literacy enriches the space of possible products that could be produced by students in our metamodel, and offers theoretical tools to interpret the collaborative and social communication processes by which the inquiry was conducted.

\subsection{Implications for Instruction}
\label{Sec:ImpliInstr}

Our metamodel allows us to identify student progress through complex, non-linear scientific inquiries that may even include subsidiary inquiries themselves. While many other computational physics courses focus on specific programming or numerical analysis skills that students may use in later courses or research \cite{graves2020hitting, weber2020benefit} or are focused on specific domains of physics problems\citep{Rebbi.2008}, our approach instead focuses on students' developing a ``grasp of practice''\cite{ford2008grasp} of computational modeling in physics. The design of our course connects to and builds upon other work on computational modeling, including computational modeling introductory courses\cite{Chabay.2008,caballero2012implementing,irving2017p3} while also drawing on research in  K-12 STEM education \cite{ford2008grasp, mcelhaney2020using,weigel2016predicting, gravel2017integrating}. Those developing college level computational physics courses may benefit from exploring curricula and methods from these other sources in addition to integrating making projects into the curricula.
 
We see in our learning environment that supports students' agency over what to model and how to model better mirrors the often chaotic practices and activity of professional scientists: the challenge is often not how to program a particular feature of a model, but rather how to make progress through a complex space of object, resources, and necessary products. 

Our metamodel provides a convenient tool grounded in philosophy of science to understand, provide feedback on and potentially assess student practice and work. In  particular, the metamodel may allow a suitably trained instructor to identify key moments---which may happen infrequently and otherwise be challenging to spot---where students make epistemic decisions around what type of knowledge to create. The metamodel may also support instructors in identifying what practices students may need further support in while engaging in computation. As we observe in the above analysis, some practices emerge more naturally than others and strategic interventions by the instructor may yield considerable improvements. For example as students assess the quality of evidence available from their work, the instructor could notice and call students' attention to the standards of evidence at stake. They could provide further resources for quantitative measures of evidence, estimating uncertainties etc.

Presenting students with the metamodel may help students' understand their own progress as well as providing a tool to support metacognition \cite{gok2010general} and reflective practice \cite{taylor2008teaching} in their computational work. Exploring training strategies for instructors and students on how to use the metamodel and how they actually use it are targets of our future work. 

\subsection{Implications for research and open questions} 

Our metamodel in this work was used to design projects and inform a content analysis of student work from computational making projects. As the metamodel provides a lens for student engagement, a next step in this work is to use the metamodel to inform video analysis of students engaging this work. By analyzing written work, we can identify what was done, but not by whom and how students navigated their pathways through activities as a group. Furthermore, we would like to understand how students and groups interact with each other--in this course, groups regularly presented intermediate work to their classmates. Video interaction analysis\cite{jordan1995interaction} and analysis of written work all groups in a class--not just the three presented here---would provide further insight. Such work would also provide the opportunity to study the role of the instructor in facilitating students' computational work and provide further recommendations for instruction.
 
While our metamodel is a valuable tool for designing curricula and understanding student practice, we do not believe the practices, artifacts, and knowledge listed in our metamodel are (or need be) comprehensive. Further examination of the computational practices of professional physicists and students will likely expand our view further. We anticipate that emerging computational methods from data science and machine learning methods can be readily accommodated in our framework, because although these part from traditional physics uses of computing, they nonetheless contribute to the overall act of scientific inquiry. 
 
We therefore believe the overarching components of the metamodel: resources, objectives, products, production, and critique will provide a valuable framework for conducting further research into other computational physics learning environments such as $P^3$\cite{irving2017p3} and C2STEM\cite{hutchins2020c2stem}. In particular, taking expansive views of both resources and products may help identify the diverse ways in which students may be ``productive'' in their work. That productivity may consist of developing accurate computational models or a deeper understanding of the physics content. It may also consist of engaging in the practices of physics and computation as productive disciplinary engagement \cite{agarwal2019integrating,engle2002guiding}, even if students fail to produce functional code \cite{kapur2008productive,kapur2016examining}.

By positioning students' intuition, lived experience, and even identity as resources \cite{nasir2009becoming}, we can design making projects that position students from all backgrounds as developing experts, similar to Basu et al's description of \emph{critical science agency} \cite{basu2009developing}. Furthermore, by asking students to present a range of different products, we more closely mirror scientific communication. Indeed, we note that our work is emphasizes communication in a similar way to recent works on Computational Literacy\citep{Odden.2020993,Odden.2019}.

Finally, we see this work as an important step in designing physics classroom spaces that center students' agency. While Physics Education Researchers have studied the role of agency---albeit in a limited way---in instructional laboratories \cite{phillips2021not,holmes2020developing}, less attention has been paid to how to design other types of classes around student agency. Even efforts on lab reform might benefit from a clear metamodel for experimental inquiry such as provided by extensions of our metamodel. Furthermore, the metamodel may help us understand the iterative, non-linear nature of students concerted activities \cite{Kelly2016}, another area that has been understudied in PER. 

\section{Conclusion}\label{sec:Conclusion}

In this work, we have presented a metamodel of computational modeling. This metamodel is grounded in philosophy of science, particularly the work of Staley \citep{staley2004robust,staley2012strategies,staley2020securing} and inspired by Humphreys's careful analysis of computational models \cite{humphreys2004extending}. It is also grounded in professional practices of computational physicists which have been examined by a number of authors\cite{Burke.2017, behringer2017aapt, Caballero.2015}. 

Our five overarching components within the metamodel are Objectives, Resources, Products, Production, and Critique. Within each of those components are categories of practices (in Production and Critique), artifacts (in Products and Resources), and abstraction and knowledge (in Products, Resources, and Objectives). We then described a computational physics course designed with the principles of agency and computational making and used our metamodel to analyze student work from that course. 

Our metamodel provides an invaluable lens for understanding student engagement and practice in computational physics, both as a goal in and of itself and for identifying how students' computational practices relate to those of professional physicists more broadly. Furthermore, the metamodel provides insight into how to go about refining our design and incorporating computation into other learning environments: it is crucial to value the inevitably non-linear and idiosyncratic nature of student work as this mirrors the practice of scientists more generally. In exerting agency over their pragmatic and epistemic objectives, we see students engaging in a wide range of production and critique practices. Rather than measuring student learning as their ability to reach pre-determined objectives or master particular skills, the metamodel gives us a framework for valuing student progress as developing a grasp of computational practice \cite{ford2008grasp}.

Our work should also significantly expand PER's construction of modeling in contexts outside computational modeling, because our metamodel moves beyond the representational view implicit or explicit in much prior work\citep{Brewe.2008,Hestenes.1987,Chabay.2008}. We build upon an important thread running through PER to create environments where students \emph{do physics} and where students' thinking and prior knowledge are valued\citep{hammer1996more,Hammer.2000}; our metamodel provides a valuable resource for making sense of what \emph{doing physics} might look like. 

Certain aspects of our metamodel parallel, and may further inform, analogous efforts to create metamodels of experimentation for laboratory design \citep{Zwickl.2014lgj,dounas2017student,dounas2018characterizing}. Our making environment, to be elaborated further in a subsequent paper, affords students a great deal of agency, both epistemic agency as has been a target of lab reform\citep{holmes2020developing,smith2020expectations,kozminski2014aapt,phillips2021not,dounas2017student, dounas2018characterizing, etkina2010design}, but also many other kinds of agency such as pragmatic and material agency. The very complex maneuvers theorized and described here justify and reinforce the need for deeper theoretical lenses through which we can view computational work\citep{Odden.2019,Odden.2020993} and physics more generally. 

\begin{acknowledgments}
This material is based upon work supported by the National Science Foundation under Grant Nos. DMR 1654283 and DRL 1742369. 
\end{acknowledgments}

\appendix*
\section{Instructor Prompts}

Here, we detail instructor prompts provided to the students at various
points in the project. 

A slide is projected as students walk into the room for the start
of project 3: 
\begin{quotation}
\textquotedblleft Make an oscillator\textquotedblright{}

\textquotedblleft Record with your phone\textquotedblright{}

\textquotedblleft Share photos and videos on Box or the Slack channel
for project 3\textquotedblright{} 
\end{quotation}
\noindent The instructor introduced the project as follows: 
\begin{quotation}
\emph{\textquotedblleft Can we focus our attention to...here---the
center of the room---rather than the front, for a change? We\textquoteright re
going to do something slightly different for project 3. In project
1 and 2 we approached the physical system computationally. For project
3, we\textquoteright re going to start with the physical system. And
we have on these tables we have a collection of strings, and wires,
and tape, and markers, and I guess there are cocktail sticks, and
tongue depressors, some piece of dowel, various cardboards of different
weights, bobs, playdough rolls, hoops, and cutters, cutting mats,
and rulers, and slicers\dots{} and all sorts of things we can use
to manipulate these materials. There\textquoteright s some little
pieces of plastic as well. Oh yeah, and magnets. Those are fun.''}

\emph{``The challenge for today\textquoteright s class is to make
an oscillator.''}

\emph{``You do not have to make one, you can make as many as you
want. The thing I insist on is that as you\textquoteright re making
oscillators I want you to document {[}emphasized this word{]} the
thing you make and what it does. Record what it does. With your cell
phone, you can make a video of it oscillating, or doing whatever it
does. Share your photos and videos of your oscillators as you go along.
I\textquoteright m going to bring in a couple more materials with
some more magnets that we can play around with. You can do this individually,
or in small groups. You can wonder around. You can move. You can reorganize
as you will.\textquotedblright{} }{[}Transcript from video data{]}
\end{quotation}

Throughout the project, the instructor emphasized that all of the course's projects are opportunities to develop professional practice of 
computational modeling by using what you know to explore something new (in ways that the students' chose). Students were encouraged to document their 
oscillator and develop their computational model by noticing the behavior of what they made, write down what they understood to be happening, connect
the behavior to disciplinary theories, and find places to expand their understanding. As students refined their documentation and collected data 
of their oscillator's motion, they were encouraged to iterate their design, goals, and data collection. On day 4 (of the 6 class periods), students 
delivered a one-slide presentation explaining ``What is your oscillator? What do you think is important to model? What do you actually want to get 
out of the modeling process? And, what physics is relevant to your
oscillator?'' Students finalized their projects by submitting a written 
report, their code, and delivering a short presentation of their work.

\bibliographystyle{apsrev4-2}
\bibliography{bibliography}

\end{document}